\documentclass[structabstract]{aa}

\usepackage[usenames]{color}                         
\usepackage{natbib}                         
\usepackage{graphicx}
\usepackage{amsfonts}
\usepackage{amsmath}
\usepackage{amssymb}
\usepackage{pifont}
 \usepackage{xspace}
     
\newcommand{\f}[2]{{\ensuremath{\mathchoice%
        {\dfrac{#1}{#2}}
        {\dfrac{#1}{#2}}
        {\frac{#1}{#2}}
                {\frac{#1}{#2}}
        }}}

\renewcommand{\vec}[1]{ {\mathbf #1} }

\newcommand{\jj}{ \vec j}
\newcommand{\bb}{ \vec B}

\newcommand{\U}[1]{\ensuremath{\mathrm{~#1}}}

\newcommand{\ha}{H$\alpha$}

\newcommand{\rjets}{{tension-driven upflows}\xspace}

\newcommand{\ejets}{{evaporation upflows}\xspace}
\newcommand{\ujet}{{untwisting model}\xspace}
\newcommand{\ujets}{{untwisting upflows}\xspace}

\newcommand{\bjet}{{blowout jet}\xspace}
\newcommand{\gjet}{{straight jet}\xspace}
\newcommand{\gjets}{{straight jets}\xspace}
\newcommand{\hjet}{{helical jet}\xspace}
\newcommand{\hjets}{{helical jets}\xspace}

\begin{document}

\title{A model for straight and helical solar jets:}
\subtitle{II. Parametric study of the plasma beta}
\titlerunning{Straight and helical solar jets - II. Plasma beta}

\author{E.~Pariat \inst{1}  \and  K.~Dalmasse\inst{2}   
\and C.R.~DeVore \inst{3} \and  S.K.~Antiochos\inst{3} \and J.T.~Karpen\inst{3} }

\institute{
$^{1}$ LESIA, Observatoire de Paris, PSL Research University, CNRS, Sorbonne Universit\'es, UPMC Univ. Paris 06, Univ. Paris Diderot, Sorbonne Paris Cit\'e, 5 place Jules Janssen, 92195 Meudon, France \email{etienne.pariat@obspm.fr}\\
$^{2}$ CISL/HAO, National Center for Atmospheric Research, P.O. Box 3000, Boulder, CO 80307-3000, USA \\
$^{3}$ Heliophysics Science Division, NASA Goddard Space Flight Center, Greenbelt, MD 20771, USA \\
}

\abstract
{Jets are dynamic, impulsive, well-collimated plasma events that develop at many different scales and in different layers of the solar atmosphere. }
{Jets are believed to be induced by magnetic reconnection, a process central to many astrophysical phenomena. Within the solar atmosphere, jet-like events 
develop in many different environments, e.g., in the vicinity of active regions as well as in coronal holes, and at various scales, from small 
photospheric spicules to large coronal jets. In all these events, signatures of helical structure and/or twisting/rotating motions are regularly 
observed. The present study aims to establish that a single model can generally reproduce the observed properties of these jet-like events.}
{In this study, using our state-of-the-art numerical solver ARMS, we present a parametric study of 
a numerical tridimensional magnetohydrodynamic (MHD) model of solar jet-like events. Within the MHD paradigm, we study the impact of varying the atmospheric plasma $\beta$ on the generation and properties of solar-like jets.}
{The parametric study validates our model of jets for plasma $\beta$ ranging from $10^{-3}$ to $1$, typical of the different layers and magnetic environments of the solar atmosphere. Our model of jets can robustly explain the generation of helical solar jet-like events at various $\beta \le 1$.  This study introduces the new result that the plasma $\beta$ modifies the morphology of the helical jet, explaining the different observed shapes of jets at different scales and in different layers of the solar atmosphere.}
{Our results allow us to understand the energisation, triggering, and driving processes of jet-like events. Our model allows us to make predictions of the impulsiveness and energetics of jets as determined by the surrounding environment, as well as the morphological properties of the resulting jets.}
  
 
 \maketitle
 
\color{black}
 
\section{Introduction}  \label{sec:Introduction} 

In the solar atmosphere, jet-like structures, defined by an 
impulsive evolution of a thin collimated bright or dark structure extending along a particular direction, are ubiquitous.
Jet-like events are observed in a wide range of environments, on 
scales ranging from the limit of instrumental resolution to hundreds of Mm. They 
have been detected in almost all wavelengths available to observers, and have thus acquired a multitude of names: 
spicules \citep[e.g.,][]{Sterling00,DePontieu07};  photospheric jets \citep[e.g.,][]{Nishizuka08,Nishizuka11};
chromospheric \ha\ surges \citep[e.g.,][]{Schmieder95}; chromospheric Ca II H jets \citep[e.g.,][]{Morita10}; 
coronal EUV jets and macrospicules \citep[e.g.,][]{Nistico09,Kamio10}; 
coronal X-ray jets \citep[e.g.,][]{Savcheva07}; and white-light polar jets \citep[e.g.,][]{WangYM02}.  
Multi-wavelength observations show slightly different 
spatial, physical, and temporal properties in each observational bandwidth, revealing that each 
jet event is formed of multi-thermal and multi-velocity plasmas
 \citep[e.g.,][]{Chifor08b,Madjarska11,Tian14}. 

In addition to the basic morphological properties that allows all these events to qualify as ``jets,'' 
 several other features and properties have been observed in some particular events independently of the scale. 
Recent high-resolution observations have shown that the base of some chromospheric jets \citep{LiuW11,ZhinchengZ16} have the same 
morphology and presence of bi-directional flows seen in examples of 
larger-scale coronal jets \citep{Shibata92}. This suggests that they may share a common underlying topological structure: a 3D null point and its 
fan/spine separatrices \citep{MorenoInsertis08,LiuC11,ZhangQM12,ZhinchengZ16}.  Regarding the morphology of the spire, using Hinode data, \citet{Suematsu08} and latter
\citet{Sterling10a} have noted that spicules present a double-stranded morphology that is similar to the emission pattern observed in coronal jets. Supersonic, though possibly sub-Alfv\'enic velocities have also been noted for chromospheric 
events \citep{Tian14} as well as jets developing in the corona \citep{Cirtain07}. Jet-like events also tend to recur homologously 
at the same location independently of their size; this property 
has been commonly observed for spicules \citep{DePontieu07}, chromospheric jets \citep{Tian14}, and 
coronal jets \citep{JiangY07,Chifor08a,Chifor08b}. 

One of the most puzzling properties that all these events display 
is the common presence of helical structure and/or twisting motion.  Signatures of rotating structure are 
present in a noticeable sample of all these phenomena in spite of their very different environments.
On the larger end  of the spatial scale, the existence of helical properties has been very commonly noticed with a broad range of techniques. 
First, the morphology of the coronal jets has been noted from X-ray and EUV images by various studies 
\citep[e.g.,][]{Shibata92,Canfield96,JiangY07,ShenY12}. The frequency of helical structure in coronal jets is strongly dependent 
on the wavelength of observation. By looking only at X-ray images, \citet{Shimojo96} reported a 10\% occurrence of twisting within their observational 
sample. Using more recent Hinode/XRT observations, \citet{Savcheva09b} noted that 14\% of their X-ray jets showed signs of twisting. In contrast, using EUV 
observations from the STEREO mission \citep{Kaiser08}, \citet{Nistico09} found that 31 out of 79 jet events presented a clear helical structure. 
From a sample of 54 jets observed in X-rays, \citet{Moore13} found that 29 out of the 32 jets that also presented a cool component of 
emission at SDO/AIA 304 \AA\ displayed a rotation 
about its axis in that channel. Helical structure and twisting motion are thus predominantly noted in the cooler EUV lines such as 304 \AA.
Using that particular bandwidth, several studies \citep[e.g.,][]{ShenY11,ChenHD12,HongJC13} have inferred the twisting rate and the helical morphology 
by analyzing the motion of well identified features within the jet structure. 

The helical shape and twisting have been further demonstrated by more advanced methods. Exploiting the stereoscopic capabilities of the joint 
observations by the two STEREO spacecrafts, \citet{Patsourakos08} reconstructed a coronal jet in 3D and unambiguously 
revealed its helical shape. Doppler images have also been used to show that several coronal jet events presented strong rotational motions, 
blue- and red-shifts being observed on opposite sides of each jet \citep{Dere89,Pike98,Harrison01,Jibben04,Cheung15}. 
Combined spectroscopic and stereoscopic observations of a jet have been carried out by \citet{Kamio10}, who also recovered its helical structure 
and confirmed its rotating nature during the event.
 
Following the conceptual ideas of \citet{Shibata85}, \citet{Shibata86}, \citet{Canfield96}, and \citet{Jibben04}, 
in recent years several numerical works have suggested and shown evidence that \hjets were driven/accelerated 
at least partly due to propagating nonlinear Alfv\'enic waves \citep{Torok09,Pariat09a,Pariat10,Pariat15a,Lee15}. 
As discussed in Sect. 2 of \citet[][hereafter PDD15]{Pariat15a}, three main physical processes might explain the acceleration of the plasma: the \rjets that 
directly result from the local dynamics of the reconnected field line at the reconnection site; the \ejets that are 
induced by the heat/pressure gradients in the reconnected field lines (heat created directly by the reconnection process 
or secondarily after accelerated particles have interacted with the plasma); and the \ujets that are induced by a 
global reconfiguration of the helicity within the newly opened magnetic field lines.
Magnetic reconnection between closed twisted field lines and open untwisted lines \citep[or large-scale closed surrounding field lines, as in][]{Wyper16} 
reconfigures the system 
and generates the \ujets: a nonlinear kink wave develops upward along each reconnected field 
line in order to distribute the twist along the whole extent of the field line. The compressive component of the nonlinear 
wave induces compression and heating of the plasma and creates an upward bulk flow of material. The overall 
\hjet is the result of the sequential reconnection of multiple field lines.

\citet{Torok09}, in a zero-$\beta$ simulation, showed the upward propagation of a pure torsional/kink wave that they associated 
with the jet 
(see their Fig. 4). \citet{Pariat09a,Pariat10,Pariat15a}, in $\beta=0.25$ cases, 
also argued that the helical jet consisted of \ujets driven by the propagation of torsional waves: 
these waves were induced by the sequential 
reconnection of twisted closed field lines with the straight open field. Figure 4 of \citep[][hereafter PAD09]{Pariat09a}, as well as Fig. 1 of PDD15, 
showed the dynamics of the magnetic field lines with the upward propagation of the twist along individual field lines. 
PAD09 further showed that the velocity of this propagating wave was Alfv\'enic (0.65$c_{A}$ - 0.90$c_{A}$), while the 
actual bulk plasma speed was much smaller (cf.\ their Fig. 6). Figures 10 and 11 of PAD09 showed the magnetic helicity flux, the 
Poynting flux and the fluxes of kinetic and magnetic energy at different heights, further confirming the upward propagation of energy 
and helicity at 
near-Alfv\'en speed due to the global Alfv\'enic nonlinear wave trains. In other numerical models of coronal jets generated 
in response to flux emergence, \citet{Archontis13a} also hypothesized that helical jets were driven by \ujets. 
Based on the same model, \citet{Lee15} 
found further evidence that the untwisting motion is associated with a propagating torsional wave. \citet{FangF14} also observed \ujets 
driving high-density plasma upward due to the Lorentz force associated
with the magnetic tension in the non-linear Alfven waves dominating
the divergence of the Poynting flux. They showed that thermal 
conduction was a second-order effect, only marginally enhancing the upflow of material by $2\%$.
 
 At smaller scales, evidence of rotating motions has also been deduced for chromospheric/transition region jet events. \citet{LiuW09,LiuW11} carried 
 out a thorough analysis of a jet observed in Ca II in the vicinity of an active region, and found multiple 
 signs of the untwisting dynamics of the chromospheric jet. \citet{Tian14} studied a sample of chromospheric events with the 
 IRIS instrument \citep{DePontieu14b}, and noted obvious transverse motions as well as line broadenings that they attributed 
 to the existence of twist and torsional Alfv\'en waves. At the photospheric level, there is accumulating evidence that a large fraction of 
 spicules present twisting motions \citep{Sterling10b,Sterling10a,DePontieu12}. 
 The recent IRIS data indicate that twisting/torsional motions are extremely frequent within the chromosphere and are associated with 
 spicules \citet{DePontieu14a}. At even smaller scale, beyond the resolution of imaging instruments, the
 spectrum of explosive events can also be interpreted as arising from the fast rotation of magnetic structures \citep{Curdt10,Curdt11}.

Although all these events develop in environments that exhibit substantial differences in temperature, pressure, 
plasma density, and even level of ionisation, some common characteristics are certainly shared between the jet-like structures. The idea that 
some common mechanism is triggering and/or driving some of these different classes of events has thus naturally developed 
\citep{Shibata07,Moore13,Cranmer15}. The present study builds on that idea by aiming to assess how well a model originally developed for large-scale  
coronal jets PAD09 can explain the properties of jet-like features appearing in the lower layers of the solar atmosphere. 

The 3D MHD model of PAD09 was developed to explain the helical properties of coronal jets and was found to properly match numerous features of 
a helical jet observed with STEREO \citep{Patsourakos08}. Subsequently, this model has been developed in a series of parametric studies that have 
explored different aspects of the generation of jets, such as the role of reconnection \citep{Rachmeler10}, the occurrence of homologous jets 
\citep[][hereafter PAD10]{Pariat10}, the influence of the photospheric and coronal magnetic geometry (PDD15), 
the generation of straight and helical jets (PAD10, PDD15), their propagation in spherical geometry from the gravitationally stratified solar corona into the solar wind \citep{Karpen16}.
and jets confined by closed coronal loops \citep{Wyper16,Wyper16b}.

A critical parameter that defines the different environments of the solar atmosphere is the plasma $\beta$, defined as the ratio of the gas pressure 
to the magnetic pressure. It is well established that in the highly magnetised corona, the dynamics of the plasma overall are dominated by the 
Lorentz force $(\beta <<1)$, whereas the gas pressure dominates 
within the solar interior $(\beta >>1)$. Hence, at the photospheric/chromospheric level, a transition layer exists where $\beta \sim 1$.  
The objective of the present work is to study the influence of the plasma $\beta$, in the range $[10^{-3},1]$, on the properties of the jet model of PAD09.  
By doing so, it is possible not only to study the validity of our jet model in the different layers of the solar atmosphere but also, at a given scale, to 
compare and possibly explain the  various properties of jets observed in different environments, e.g., in coronal holes or active regions.

In the highly stratified solar atmosphere, jet-like events are
expected to occur in a complex multi-$\beta$ environment.  Numerous recent numerical 
models have thus simulated different jet-like events in a stratified atmosphere including a 
rich range of physical processes: spicules \citep[e.g.,][]{MartinezSykora11,MartinezSykora13}, chromospheric jets \citep[e.g.,][]{YangL13b}, surges \citep{NobregaSivero16}, and 
coronal jets \citep[e.g.,][]{MorenoInsertis13,FangF14,Torok16}. In this paper
we complement these studies by performing simulations within a
relatively uniform atmosphere, but with varying $\beta$. By
simplifying the environment compared to the real solar atmosphere, we
are able to isolate parametrically the 
role of the plasma $\beta$ parameter in a controlled way. This facilitates clearer understanding of the 
fundamental physical processes responsible for the evolution observed in more complex numerical models.

The present study is organised as follows: 
 in Sect. \ref{sec:PaperI} we first quickly synthesise the results of the preceding study of this series \citep{Pariat15a} and in Sect. \ref{sec:Model} we summarise the main set-up of our numerical model. We then successively discuss the influence of the plasma $\beta$ on the trigger of straight and \hjets (Sect. \ref{sec:Trig}), the driving mechanism (Sect. \ref{sec:Betadriver}), and morphological properties (Sect. \ref{sec:Betamorpho}) of \hjets.  Finally, we discuss the implications of our findings in Sect. \ref{sec:Concl}.

%

\section{Summary of the results of Paper I} \label{sec:PaperI} 

\begin{figure*}[h!]
\resizebox{\hsize}{!}{\includegraphics{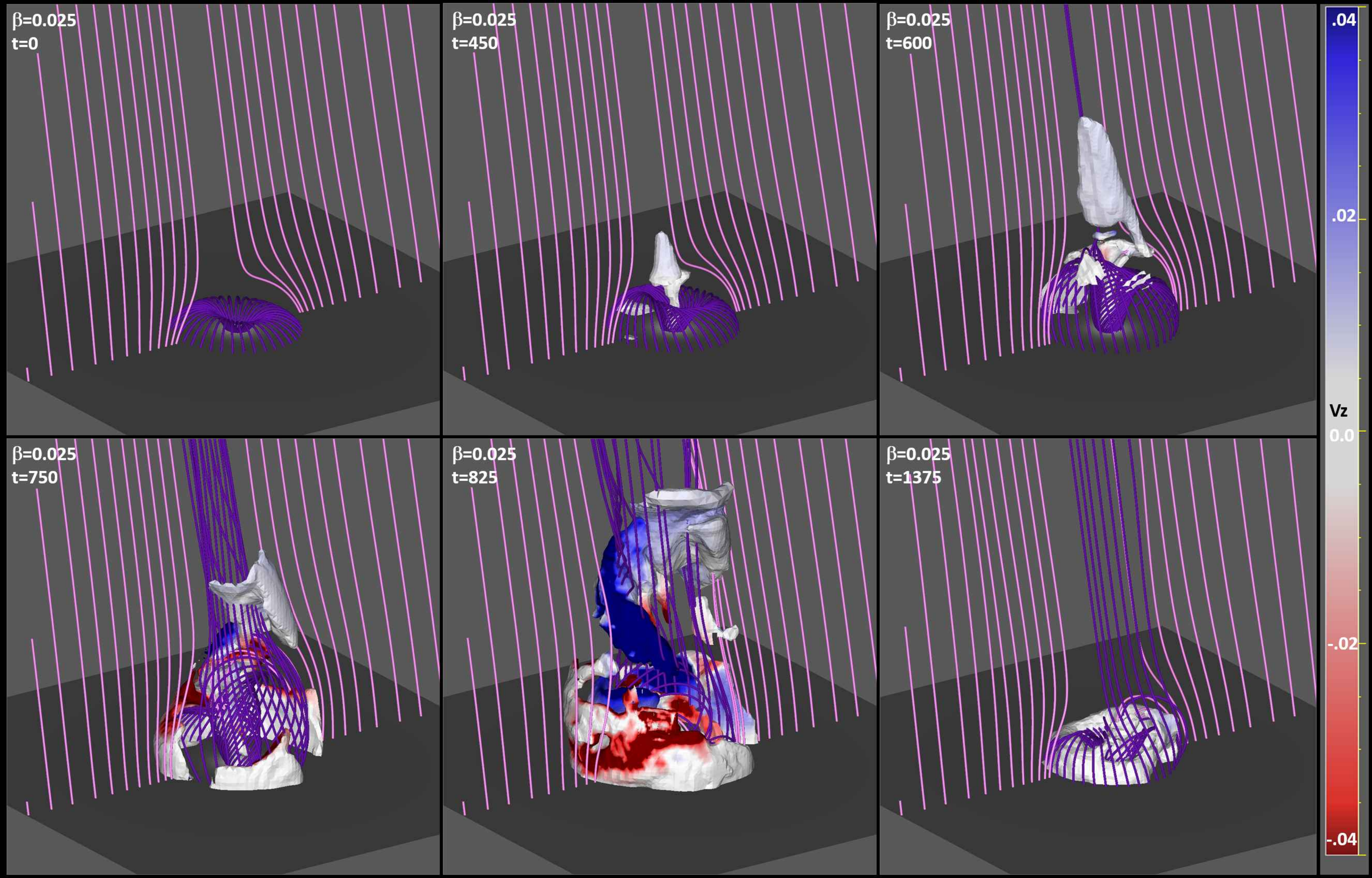}}
\caption{Snapshots of the evolution of the system during the generation of a mild straight jet (top row) and a very energetic helical jet (bottom row) for the $\beta=0.025$ simulation. The resp. [purple/pink] field lines, which are initially resp. [closed/open], are plotted from fixed points along the bottom boundary resp. [on a circle of radius 1.5$L_0$/along the y=0 axis]. The isosurface of the plasma density equal to $\hat{\rho}=1.2$ is color-coded according to the vertical velocity component $v_z$: red indicates downward velocity, and blue upward. For comparison, the initial uniform density $\hat{\rho}=1$ and the ambient Alfv\'en speed $\hat{c}_A=0.28$. The simulation domain extent is $[-12L_0;12L_0]\times[-12L_0;12L_0]\times[0;24L_0]$.
        }
       \label{FigEvolBeta0025} 
\end{figure*}

In our previous studies we introduced a model that explains morphologically different 3D coronal jets, both straight and helical. An example of the time-evolution of the system in one of the parametric simulations of this study is presented in Fig. \ref{FigEvolBeta0025}. Other examples of such dynamics with slightly different conditions can be found in Fig. 4 of \citet[][PAD09]{Pariat09a} Fig. 2 of \citet[][PDD15]{Pariat15a}, and Fig. 1 of \citet{Pariat15b}. The basic physics of our model is that a straight jet is due to slow interchange reconnection between the closed flux of an embedded bipole region \citep[e.g.,][]{Antiochos90} and the surrounding open flux of a coronal hole (note the change of connectivity of some straight pink and purple field lines in the top-middle vs. top-left panels of Fig. \ref{FigEvolBeta0025}). In contrast a helical jet is due to an explosive burst of this interchange reconnection (note the change of connectivity of a numerous strongly twisted purple field lines in the bottom-right vs. bottom-middle panels of Fig. \ref{FigEvolBeta0025}). In our model, the slow reconnection is driven by the response of the system to the continual stressing of the closed flux by photospheric motions.  On the Sun, this could also be due to continued flux emergence. The explosive reconnection is due to a kink-like instability in the closed field region when the magnetic stress builds up beyond a certain level. 

The \hjet is unleashed by the explosive interchange reconnection between open and closed magnetic fields, which generates a series of impulsive nonlinear Alfv\'enic or kink waves that propagate upward along reconnection-formed open field lines (e.g., kinked purple in the bottom-middle panel of Fig. \ref{FigEvolBeta0025}), and eject most of the twist (magnetic helicity) stored in the closed domain. The main acceleration process is explained by the \ujet of \citet{Shibata85,Shibata86}, although a tension-driven flow is embedded within the structure of the \hjet (PAD09). In Sect. \ref{sec:Betadriver}, we further study the physics and properties of the wave beyond that of our previous analysis.

In PDD15 we presented two parametric studies of the generation of \gjets and \hjets: one varied the inclination angle of the coronal magnetic field while the other varied the photospheric distribution of the magnetic field while preserving the basic topology. We confirmed that the basic model of PAD09 was valid across a broad parametric range. PDD15 showed that \hjets are triggered for inclination angles in the range $\theta = 0 - 20^\circ$. As long as a 3D magnetic null point is present, our model is also valid for different photospheric distributions of flux concentrations surrounding the central embedded-bipole polarity, configurations that are frequently observed in the solar atmosphere.

We showed in PDD15 that a \hjet was generated for all inclinations but this is not true for the \gjet. A \gjet formed only when the 3D null point was sufficiently stressed to form an extended current sheet in response to the boundary-driven motions. We found that \gjets appear only for inclination angles $\gtrsim 8^\circ$. We also found that different photospheric magnetic distributions strongly affect the generation of the \gjet . From the first parametric studies presented in PDD15 we showed that a preceding reconnection-driven straight jet profoundly influences the onset of the succeeding \hjet. The third parametric study reported here will extend these results by showing that the actual occurrence of a \gjet is not essential, the early existence of intense reconnection is the key element that affects the trigger of the helical jet.

The parametric study presented here, however, goes further than merely confirming the results of PDD15.  We present a completely new analysis of the underlying physical mechanism that accelerates the plasma in our model  (Sect. \ref{sec:Betadriver}). By varying $\beta$, we also show the original result that the morphology of the jet is strongly influenced by this parameter, a finding that was not expected from our previous simulations.

%
%

\section{Model description} \label{sec:Model} 

The simulations presented here extend the work presented in \citet[][PDD15]{Pariat15a}.
In the simulations, we consider the equations of ideal magneto--hydrodynamics (MHD) 
for a monofluid coronal plasma of uniform density and temperature. 
The simulations were performed with the 
{\it A}daptively {\it R}efined {\it M}agneto\-hy\-dro\-dynamic 
{\it S}olver ({\it ARMS}), whose Flux-Corrected Transport 
algorithms are extensions of those derived in \citet{DeVore91}.  
The time-dependent equations of ideal MHD, with the magnetic forces 
expressed in the Lorentz form, are solved on a dynamically 
solution-adaptive grid managed by the toolkit {\it PARAMESH} 
\citep{MacNeice00}. This grid refines and derefines adaptively during the simulation as prescribed in
the Appendix of \citet{Karpen12}. 
A Cartesian domain is assumed with $x$ and $y$ the horizontal axes and $z$ the vertical axis.
The simulation domain spans $[-12L_0;12L_0]\times[-12L_0;12L_0]\times[0;24L_0]$.
The nonuniform initial grid is  identical to the one presented in Fig. 1 of PAD09. 
The same boundary conditions are used as in PDD15, i.e. line-tied at the bottom, closed on the four 
sides, and open at the top.

The initial magnetic field is set to be potential, using the specific analytical form given in Sect. 3 of PDD15. 
The initial configuration  (cf. top-left panel of Fig. \ref{FigEvolBeta0025}) is 
given by a central vertical magnetic dipole placed under the photosphere (producing a locally closed coronal field, e.g., purple lines in Fig. \ref{FigEvolBeta0025}), 
embedded in an inclined (with respect to the vertical direction) and uniform 
background magnetic field (the open field, e.g., pink lines in Fig. \ref{FigEvolBeta0025}). 
A 3D null point with its associated fan surface and two 
spine lines are present, with the outer spine following the general direction 
of the open field. 
In the present parametric study, the inclination angle $\theta = 10^\circ$ is the same for all simulations. 
As in PDD15, energy is injected in the closed domain through line-tied twisting 
motions at the bottom boundary. The imposed velocity field is given by Equation 3 of PDD15.  

We follow 
the evolution of the free magnetic energy in the system, taken as
the difference, $E_{mag}$, between the total magnetic energy $E_m$ and its initial value:
$E_{mag}(t)=E_m(t)-E_m(t=0)$. This is only a proxy of the actual free magnetic energy, defined as the difference
between the total magnetic energy in the system and the magnetic energy of the potential field having the 
same distribution of the normal magnetic field across the six boundaries. Although the system is initially potential 
and $B_z$ at the bottom boundary is constant in time, since the magnetic flux distribution changes slightly on the 
other boundaries, the potential field and its energy change in time.  
\citet{Pariat15b} showed in one case 
(their Fig. 2) that the difference between $E_{mag}$ and the real free energy is at most $20\%$, and that the two values are 
monotonically related. 

The domain is filled with a highly conducting coronal plasma.  
For maximum generality, we use non-dimensional units (denoted as, e.g., $\hat{f}$). 
The initial thermal pressure, $\hat{P}$, is uniform, as is the initial 
mass density, $\hat{\rho}$. We assume an ideal plasma 
equation of state. The temperature is therefore initially uniform, $\hat{T}=\hat{P}/(\hat{\rho} \hat{R})$,
where $\hat{R}=0.01$ is the non-dimensional gas constant. The volume magnetic field $\hat{B}_v=1$ is 
used as the reference magnetic field intensity. From this non-dimensional setting, choosing the value of 
certain physical quantities (denoted, e.g., $f_0$) fully determines the physical scales of the system. The precise 
correspondence is detailed in Appendix \ref{An:Scaling}.

In this study, we analyze how modifying the plasma $\beta$ impacts the 
development of the jet.  All our previous published computations have been 
performed assuming $\beta=0.25$, defined with respect to the uniform background field strength. In the present paper we 
performed three additional simulations where the $\beta$ ranges from $1$ to $2.5 \times 10^{-3}$. 
We seek to test whether the modification of the $\beta$ value precludes the existence of the 
jet, and to establish how it modifies the jet properties and dynamics.

The plasma $\beta$ is varied by using different values for the 
non-dimensional plasma pressure $\hat{P}$.  The simulations presented in PDD15 
considered a value of $\hat{P}$ uniformly equal to $0.01$. 
We perform here additional runs with a uniform pressure in the domain, respectively equal to 
$[4\times 10^{-2}, 10^{-2}, 10^{-3}, 10^{-4}]$. The non-dimensional density is kept constant 
between the runs, $\hat{\rho}=1$, as is the volume magnetic field, $\hat{B}_v=1$. 
The plasma $\beta$ is therefore given by $\beta= 2\hat{\mu}\hat{P}/\hat{B}_v=8\pi\hat{P}$. 
The different runs correspond to the respective plasma $\beta$ values of $[1.0, 0.25, 0.025, 0.0025]$
The non-dimensional Alfv\'en velocity, $\hat{c}_A=\hat{B}_V/\sqrt{\hat{\mu} \hat{\rho}}=(4\pi)^{-1/2}\simeq0.28$ is therefore 
constant between the runs, which allows us to compare directly the
evolutions of the four cases. The non-dimensional sound speed $\hat{c}_S= (\gamma \hat{P}/\hat{\rho})^{1/2}=(5\hat{P}/3)^{1/2}$ 
decreases with $\hat{P}$ (and $\beta$), adopting the following values $[0.26,0.13,0.041,0.013]$. The parameters used in the four simulations
are summarized in Table \ref{Tab:ScalNonDim}.

\begin{table}[h!]
\caption{Characteristic of different simulations in non-dimensional units: $\beta$, pressure $\hat{P}$, Alfv\'en speed  $\hat{c}_A$,  sound speed $\hat{c}_S$.
}
\label{Tab:ScalNonDim} 
\centering
\begin{tabular}{cccc}
\hline 
$\beta$ & $\hat{P}$ & $\hat{c}_A$ & $\hat{c}_S$ \\
\hline 
$ 1.0 $ & $ 0.04 $ & $ 0.28 $ & $ 0.26 $  \\ 
$ 0.25 $ & $ 0.01 $ & $ 0.28 $ & $ 0.13 $   \\
$ 0.025 $ & $10^{-3} $ & $ 0.28 $ & $ 0.041 $ \\ 
$ 0.0025 $ & $10^{-4} $ & $ 0.28 $ & $ 0.013 $\\ 
\hline
\end{tabular}
\end{table}

These four simulations allow us to simulate a wide variety of conditions in the solar atmosphere. 
The tables presented in Appendix \ref{An:Scaling} present various possible physical scalings that correspond to our different cases. 
The simulations using different values of $\beta$ allow us more particularly to determine the dynamics of jets occurring in the 
different layers of the solar atmosphere. Table \ref{Tab:Scaling} presents one set of dimensional quantities 
that corresponds to different layers of the solar atmosphere for each run. 
The $\beta=0.025$ and  $\beta=0.0025$ runs correspond to corona-like conditions, whereas the transition region
would be typically represented with $\beta=0.25$. The $\beta=1$ run is applicable to some aspects of photosphere/chromosphere-like conditions. This correspondence is suggestive, but not definitive, since 
the chromosphere is a layer with strong density stratification while our simulations assume uniform ambient density. Due to the non-dimensional nature of the MHD system solved here, different $\beta$ simulations can actually correspond to a wide range of parameters. As is shown in the Appendix \ref{An:Scaling}, a given $\beta$ simulation can apply 
to different layers for different values of the magnetic field, $B_0$. The layer correspondence in Table \ref{Tab:Scaling} is only one possibility.
 
Note that our simulations focus on calculating the magnetically-driven
dynamics of solar jets and not on the plasma thermal properties;
consequently, we use a simple adiabatic energy equation and numerical
resistivity. Our simulations do not include effects such as thermal
conduction, or radiative transfer, or a generalized Ohm's law with Hall
and ambipolar-diffusion. These effects are well known to be important
in the real corona and chromosphere for determining the plasma
energetics \citep[see e.g.,][]{MartinezSykora12,Leake13a,Leake14b}. 
Additionally, our simulations assume strict line-tying conditions at the lower boundary. 
The primary challenge in numerical simulation of solar jets, however is
not the plasma thermodynamics, but the effective Lundquist number of
the simulation. Even in the chromosphere the Lundquist number is high,
$> 10^6$ or so, well beyond the reach of present 3D simulations. It is
absolutely essential that the numerical Lundquist number be as high as
possible so that the resulting evolution is determined by true
helicity-conserving reconnection rather than by simple diffusion. Our
adaptive mesh refinement code {\it ARMS} does an excellent job at
conserving helicity, even for reconnection-dominated evolutions
\citep[e.g.,][]{Knizhnik15,Pariat15b}.

\begin{table*}[h!]
\caption{Possible scaling for the simulations: characteristic length $L_0$ (in Mm), time $t_0$ (in s), volume magnetic field $B_0$ (in G), pressure $P_0$ (in Pa), density $\rho_0$ (in$\U{kg\ m^{-3}}$), temperature $T_0$ (in K), velocity $V_0$ (in$\U{km\ s^{-1}}$), Alfv\'en speed $c_A$ (in$\U{km\ s^{-1}}$), sound speed $c_S$ (in$\U{km\ s^{-1}}$), and energy $E_0$ (in J). The last column suggests an equivalence with different layers of the solar atmosphere.
}
\label{Tab:Scaling} 
\centering
\begin{tabular}{cccccccccccc}
\hline 
$\beta$ & $L_0 $ & $t_0$ & $B_0$ & $P_0$ & $\rho_0$ & $T_0$ & $V_0$ & $c_A$ & $c_S$ & $E_0$ & Solar layer\\
\hline 
$ 1.0 $ & $ 0.05 $ & $ 1.1 $ & $ 3 $ & $ 0.04 $ & $ 4.4 \times 10^{-10} $ & $ 10^{4} $ & $ 45 $ & $ 12 $ & $ 11 $ & $ 1.1 \times 10^{14} $ & Chromosphere \\ 
$ 0.25 $ & $ 0.8 $ & $ 1.4 $ & $ 3.2 $ & $ 0.01 $ & $ 3.1 \times 10^{-12} $ & $ 4.0 \times 10^{5} $ & $ 574 $ & $ 162 $ & $ 74 $ & $ 5.2 \times 10^{17} $ & Transition Region \\
$ 0.025 $ & $ 5 $ & $ 1.7 $ & $ 5 $ & $ 0.0025 $ & $ 3.0 \times 10^{-13} $ & $ 10^{6} $ & $ 2873 $ & $ 810 $ & $ 117 $ & $ 3.1 \times 10^{20} $ & Corona\\ 
$ 0.0025 $ & $ 25 $ & $ 1.9 $ & $ 10 $ & $ 10^{-3} $ & $ 6.1 \times 10^{-14} $ & $ 2.0 \times 10^{6} $ & $ 1.3 \times 10^{4} $ & $ 3624 $ & $ 165 $ & $ 1.6 \times 10^{23} $& Corona\\ 
\hline 
\end{tabular}
\end{table*}

%
%

\section{Influence of the plasma beta on the trigger of straight and helical jets} \label{sec:Trig} 

\begin{figure*}[h!]
\sidecaption
\resizebox{\hsize}{!}{\includegraphics{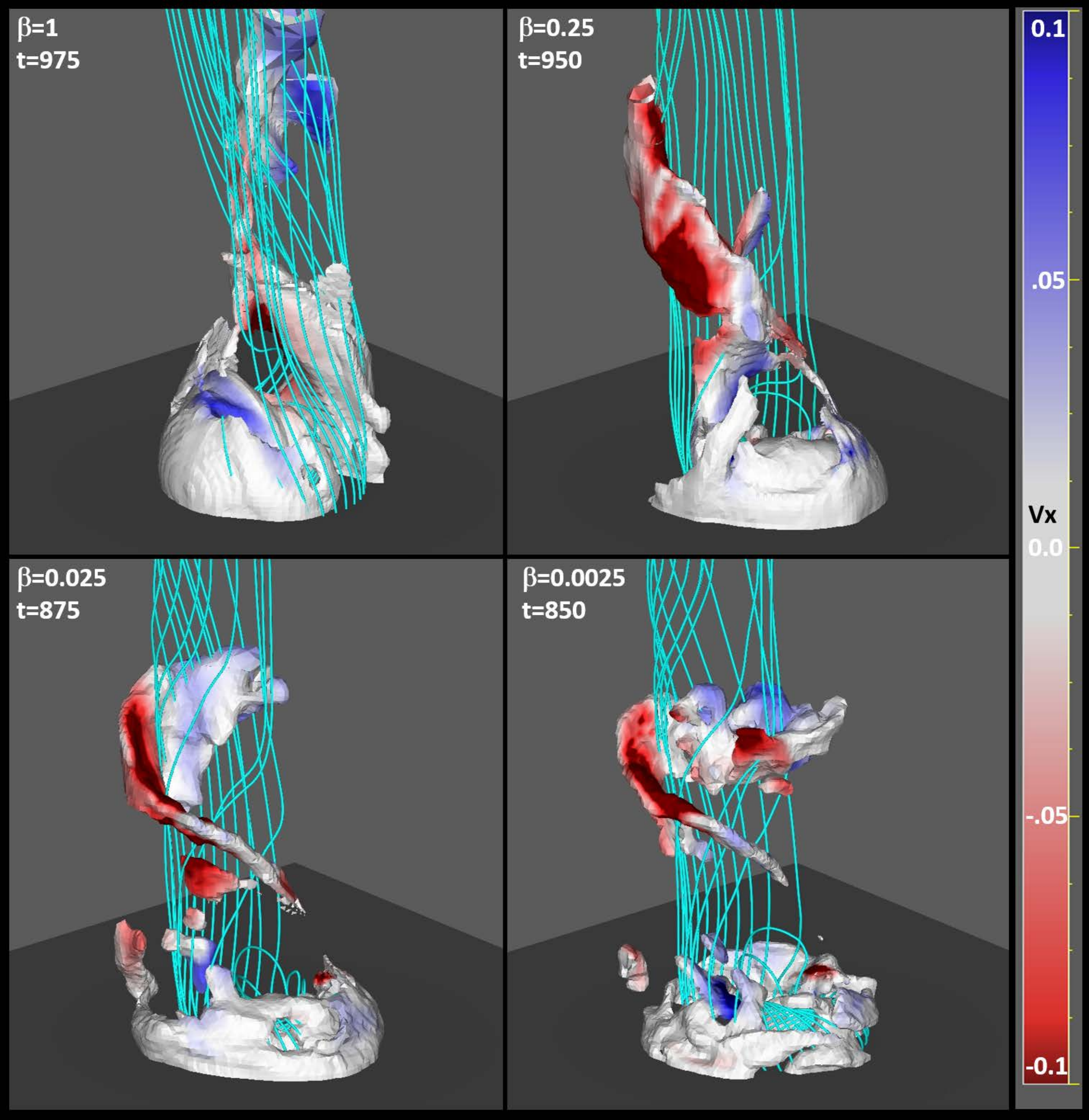}}
\caption{Morphology of the helical jet during the blowout for simulations with different plasma $\beta$. The cyan field lines, which were all initially closed, are plotted from fixed points along the bottom boundary on a circle of radius 1.6$L_0$. The helical jet in each simulation is highlighted by an isosurface of the plasma density equal to $\hat{\rho}=[1.05; 1.2; 1.6; 2.3]$ from the highest to the lowest $\beta$ case respectively, color-coded according to the transverse velocity component $v_x$: red indicates velocity oriented toward the right, and blue toward the left. For comparison, the initial uniform density $\hat{\rho}=1$ and the ambient Alfv\'en speed $\hat{c}_A=0.28$. At the centre of the domain the field of view extends vertically from $0$ to $\approx 20L_0$.
        }
        \label{Fig:BetaDyn} 
\end{figure*}

\begin{figure*}[h!]
\sidecaption
\resizebox{\hsize}{!}{\includegraphics{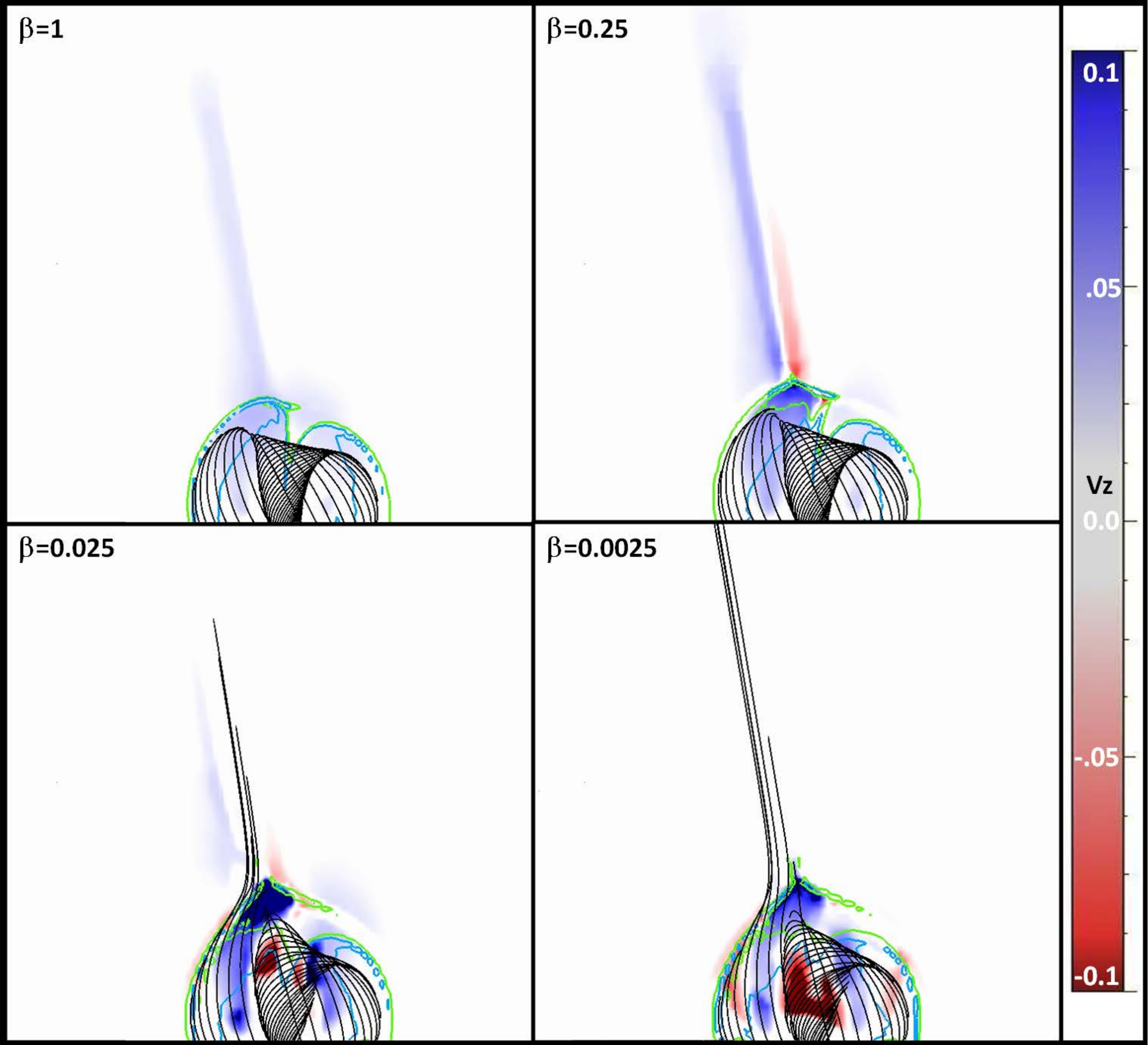}}
\caption{Vertical velocity, $v_z$ distribution in the $y-z$ plane at $x=0$ at $t=625$, during the \gjet/pre-\hjet phase for simulations with different plasma 
$\beta$ highlighting the presence of a \gjet for the runs at higher $\beta$.The velocity magnitude is color coded in blue (upflows) and red (downflows).  For comparison, the ambient Alfv\'en speed is $\hat{c}_A=0.28$.
The black field lines, all initially closed correspond to the cyan field lines of Fig.~ \ref{Fig:BetaDyn}. Open black field lines for lower 
$\beta$ runs indicate the occurrence of relatively more intense reconnection by this stage. The blue and green lines are isocontours of the electric 
current density. 
        }
        \label{Fig:BetaStand} 
\end{figure*}

\begin{figure}[h!]
\resizebox{\hsize}{!}{\includegraphics[width=\hsize]{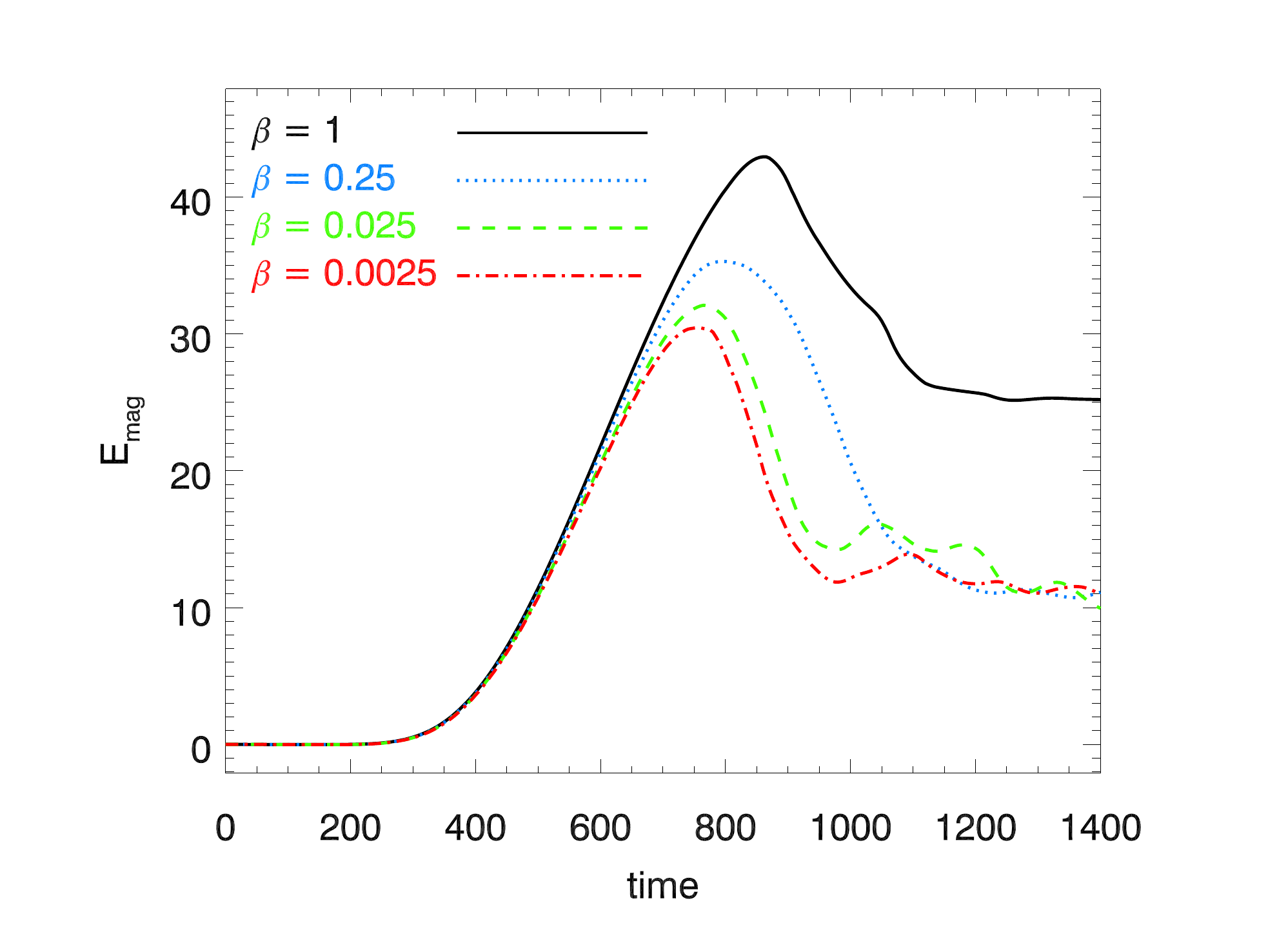}}
\resizebox{\hsize}{!}{\includegraphics[width=\hsize]{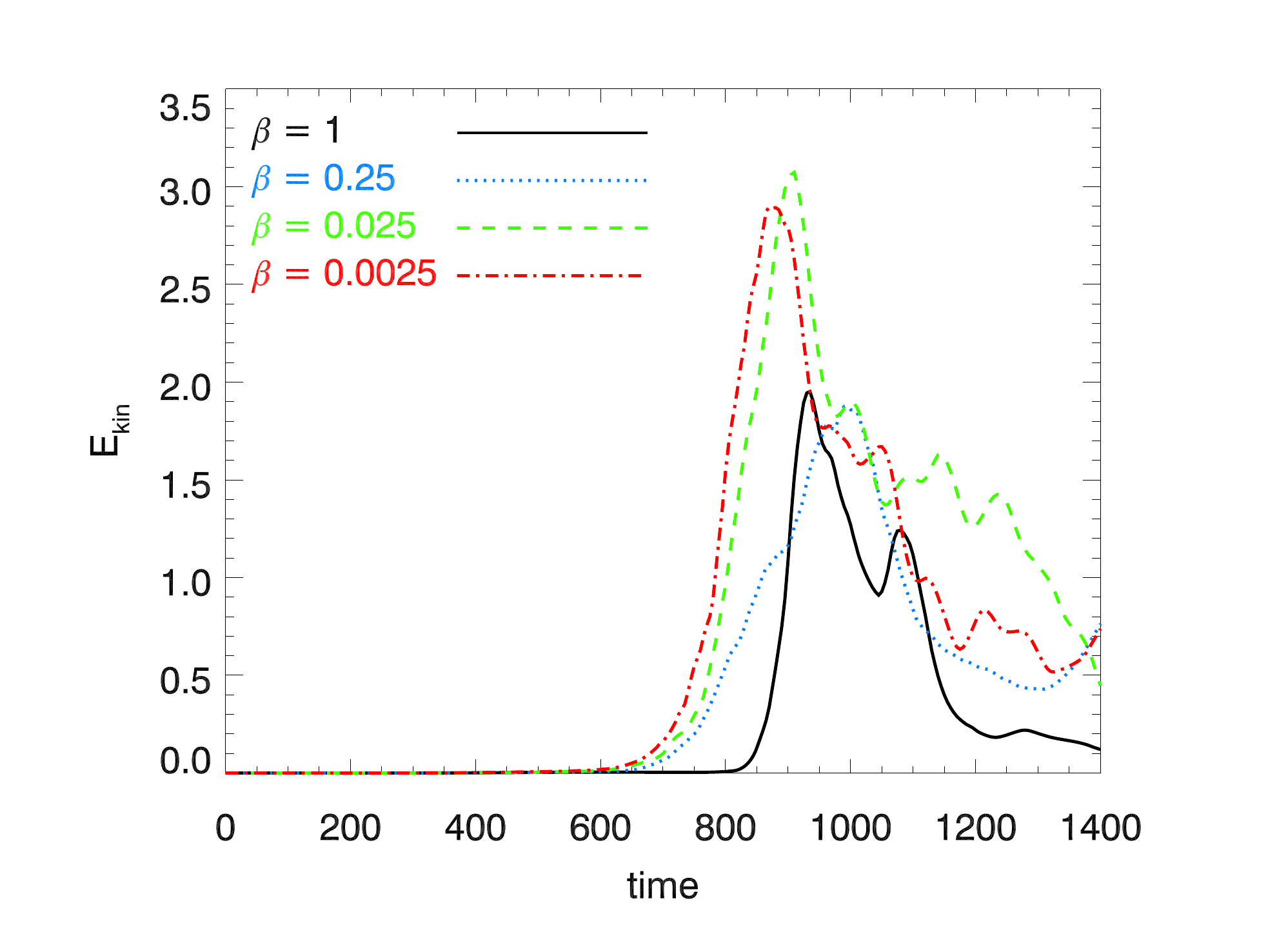}}
\caption{Top panel: time evolution of the free magnetic energy, $E_{mag}$, for simulations with different plasma $\beta$. Bottom panel: time evolution 
of the kinetic energy, $E_{kin}$, in each simulation.
        }
        \label{Fig:BetaE} 
\end{figure}

In this section we examine how the plasma $\beta$ influences the generation of the \gjet and \hjet.
The morphology of the \hjet for each run is presented in Fig. \ref{Fig:BetaDyn}, 
while the dynamical state of the system at $t/t_0=625$, 
at the onset of the \gjet phase, is presented in Fig. \ref{Fig:BetaStand}. 
The evolution of the free magnetic energy, $E_{mag}$ (defined 
earlier in Sect. \ref{sec:Model}), and the kinetic energy, $E_{kin}$, for 
the runs at different $\beta$ are presented in Fig. \ref{Fig:BetaE}. 

The first result of this parametric study is that, for all the values of $\beta$ tested, a \hjet 
always occurs eventually. Figure \ref{Fig:BetaDyn} shows the existence of a higher-density/higher-temperature region related 
to helical upflows and the presence of an upward propagating nonlinear Alfv\'enic wave, as first introduced in PAD09 and discussed 
further in Sect. \ref{sec:Betadriver} below. 

We observe in all runs that the jet consists of a left-handed helical density structure. 
The specific density distribution (shown by the isodensity surfaces) in the different \hjets exhibits variations in the 
width and pitch of the helical density structure, which we discuss further in Sect.\ref{sec:Betamorpho}. 
All of the simulations present a similar distribution of $v_x$ (red/blue color-coding of the isodensity surface) indicating 
plasma flowing away from the observer on the left side of the jet and toward the observer on the right. For each parametric run, 
this red/blue pattern on opposite sides of the jet characterises the strong clockwise rotation associated with the untwisting of the reconnected field lines. 

The free magnetic and kinetic energy curves shown in Fig. \ref{Fig:BetaE} all follow the typical 
evolution observed previously (PDD15) during the 
generation of the \hjet. There is a peak of the 
free magnetic energy, followed by a sudden drop, corresponding to the release of 
magnetic energy by intense magnetic reconnection (top panel). The 
partial transformation of magnetic to kinetic energy results in a sharp peak in the latter (bottom panel). 
The changes in the free energy are proportional 
to the intensities of the kinetic energy, as shown in PAD10. Quantitative differences are observed among the different runs regarding the \hjet 
trigger, however: the trigger time and the free energy levels differ from one simulation to another. 
Using Equation 5 of PDD15, we derive the trigger time, $T_{trig}$, and the 
trigger energy, $E_{trig}$, from the peak of the free magnetic energy curves (Fig. \ref{Fig:BetaE}, 
top panel). The resulting values are given in Table \ref{Tab:BetaParam}. We find consistently that for increasing 
values of $\beta$, the \hjet tends to occur later, after a larger amount of energy has been stored.

In PDD15, we observed that the \gjet developed before the onset of the \hjet for sufficiently large values 
of the inclination angle ($\theta > 8^\circ$).
The present $\beta=0.25$ run (with $\theta > 8^\circ$) hence also develops a \gjet with a 
higher-density region and marked upflowing vertical 
velocities aligned along the outer spine (upper right panel of Fig. \ref{Fig:BetaStand}). 
However, no similar \gjet is observed for the runs 
with lower values of $\beta$ before the onset of the \hjet. For example, only an small upflow with a relatively limited 
vertical extent is observed  in the $\beta=0.025$ run (lower left panel 
of Fig. \ref{Fig:BetaStand}). Similar to the case $\beta=0.25$, the \gjet is present for the $\beta=1$ simulation. 
In the previous parametric simulations (cf. Fig. 4 of PDD15), we noted  that the \gjet 
preceding the formation of the \hjet also was accompanied by an increase of the kinetic energy. Fig. \ref{Fig:BetaE}, lower panel, displays a similar behaviour (which was not observed in he axisymmetric case of PAD09). 

The absence of a \gjet for the lower $\beta$ runs does not mean that reconnection is 
absent in this phase. On the contrary, reconnection actually 
occurs in the lower $\beta$ runs, as well. This can be noted first in Fig. \ref{Fig:BetaStand} 
by the existence of longer and more intense current sheets: 
for smaller $\beta$, the length of the blue and green isocontours of the current 
density in the vicinity of the null point increases. In addition, for smaller $\beta$, one observes that some black field lines are open, 
whereas they initially belonged to the closed domain. Furthermore, we observe in Fig. \ref{Fig:BetaE} that the lower-$\beta$ curves of the 
free magnetic energy present a slightly more gentle slope, an indication of the magnetic dissipation occurring at the reconnection site. 
This demonstrates that relatively more reconnection has occurred at $t/t_0=625$ for the runs with smaller $\beta$, even though the driving is the same. 

By analyzing the moment at which the intense current sheet appears (not shown here), we find that the lower the $\beta$, the earlier 
the current sheet forms at the 3D null point, and the more intense is the reconnection when it develops. This indicates that the lower the plasma 
$\beta$, the earlier and the stronger is the reconnection during the \gjet phase.  However, in our low-$\beta$ simulations, this more intense
reconnection does not result in the formation of a more marked high-density region and extended upflows that would 
be interpreted as an \gjet.  In the high-$\beta$ regime, reconnection may more efficiently produce high-density upflows, 
while at lower $\beta$, although more magnetic energy is released by reconnection, this energy does not drive 
high-density plasma flows.  

A possible physical origin of this result may be related
to our assumed initial conditions. Note that the initial state has a
static, uniform density and temperature plasma and a potential
magnetic field. Consequently, if interchange reconnection were to
occur very easily, for example as soon as the photospheric driving is
turned on, no straight jet would be observed, because the closed
plasma released by the reconnection would have identical properties to
the open. Furthermore, the nonlinear Alfv\'en wave flux produced by the
reconnection would be minimal. In order for the interchange
reconnection to produce a large effect, the closed flux must undergo a
substantial deformation with a large current sheet built up at the
deformed null and a compression of the closed plasma against the
dome-like separatrix. This result re-emphasizes the point raised above
on the critical importance of the effective Lundquist number of the
simulation. We find that, for very low plasma $\beta$, interchange
reconnection begins so easily that the released closed plasma is near
its initial state. It should be noted, however, that in the real
corona the plasma in closed field embedded bipoles generally has
higher temperature and density than surrounding open field plasma. In
this case any interchange reconnection, even for very low $\beta$,
would result in an observable straight jet. However, our basic result is
still valid, the higher the $\beta$ the higher the density of the
released plasma (compared to its initial state).

As noted earlier, for higher $\beta$, the \hjet is generated later and after a larger amount of energy 
has been stored. Thus, there is a correlation 
between the low values of $\beta$, the large reconnection in the \gjet phase (whether or not a \gjet is actually present), 
and the earlier and lower energy 
trigger of the \hjet. In PDD15, our parametric studies of the field inclination and the photospheric flux distribution 
demonstrated that more intense reconnection in the pre-helical-jet phase (e.g., during the straight-jet phase) 
was correlated with an easier trigger of the helical jet, at lower energy and earlier in time. 
This third parametric study of the plasma $\beta$ further confirms our previous findings: 
the reconnection during the \gjet phase strongly influences the timing and energetics of the \hjet phase. Our current results 
suggest, with respect to our other parametric study, that it is not the plasma $\beta$ that directly triggers the \hjet at a lower energy threshold, but
rather the intermediate action of reconnection developing during the \gjet phase, prior to the \hjet onset.  
A stronger current sheet at the null and more intense reconnection during the \gjet phase trigger the helical jet earlier and 
at a lower stored-energy level.

Why reconnection is stronger at lower $\beta$ can be partly understood by the larger growth of the 
closed domain for the lower-$\beta$ runs. As discussed in PDD15, the reconnection at the 3D null during the \gjet phase 
is driven by the magnetic forcing imposed at the bottom boundary.  While the amount and rate of energy injection is exactly the same in all runs, 
the plasma reacts differently to the driving depending on $\beta$. Indeed, the increase in the magnetic pressure imposed by the twisting 
is more strongly compensated by the plasma pressure at higher $\beta$.  The higher the $\beta$, the less the closed-field 
domain bulges as a consequence of the twisting. This can be observed in Fig. \ref{Fig:BetaStand}, 
where the closed domain (distinguished by the black field lines) occupies a larger volume for lower value of $\beta$. The growth of the closed 
domain induces a stress on the 3D magnetic null point. The more the closed domain bulges, the more 
stressed is the 3D magnetic null point, and the faster reconnection develops. This explains why, during the \gjet phase,
 the reconnection is stronger for lower values of $\beta$, as the magnetic reconnection is more easily 
 forced. Then, as the reconnection is stronger for lower values of $\beta$, it leads to an easier \hjet 
generation, as seen in previous parametric simulations. 

Overall, the trigger and driver of the \hjet are similar in all runs, while the intensity of the \gjet 
depends sensitively on the value of $\beta$.
The coronal jet model of PAD09, PAD10, and PDD15 generates helical jets for a wide range of plasma $\beta$. 
The helical jets for $\beta$ of magnitude $~10^{-2}$ and $~10^{-3}$ are very similar. Hence, we argue that 
our jet model would also act similarly at still smaller values of $\beta$, as the influence of the plasma 
pressure becomes even more negligible. Overall, therefore, the model is able to produce multiple types of 
helical jet-like events that are observed to occur in 
the various layers of the solar atmosphere (cf.\ Sects. \ref{sec:Introduction} and \ref{sec:Concl}).

%
%
\begin{figure*}[h!]
\resizebox{0.9\hsize}{!}{\includegraphics[height=\hsize]{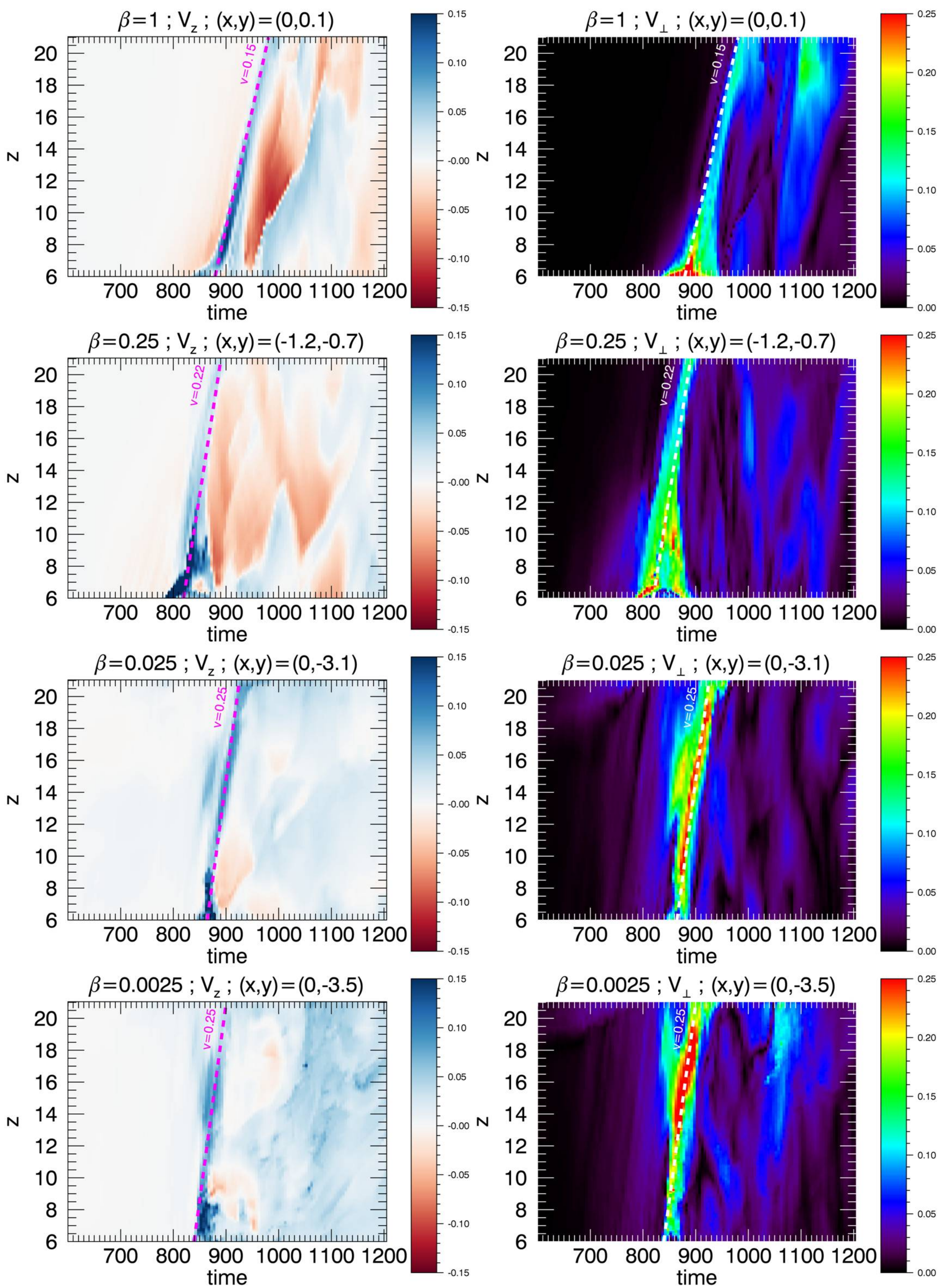}}
\caption{Time evolution of the vertical and horizontal velocities (respectively $v_z$ and $v_\perp$) along the vertical direction $z$ at 
particular points $(x,y)$, in units of $L_0$, in the four simulations at different $\beta$. The phase velocity $v_\varphi$ of the propagating wave is indicated by dashed lines. For comparison, the ambient Alfv\'en speed is $\hat{c}_A=0.28$.
        }
        \label{Fig:TsliceVelocity} 
\end{figure*}

\section{Driver of the helical jet} \label{sec:Betadriver} 

As discussed in the Introduction, previous studies have suggested that \hjets were driven/accelerated by \ujets thanks to propagating 
nonlinear waves. In the present section, we further study how the \ujet can drive the plasma upward and form the helical structure of the jet in different $\beta$ environments.

\begin{figure*}[h]
\resizebox{0.95\hsize}{!}{\includegraphics[width=\linewidth]{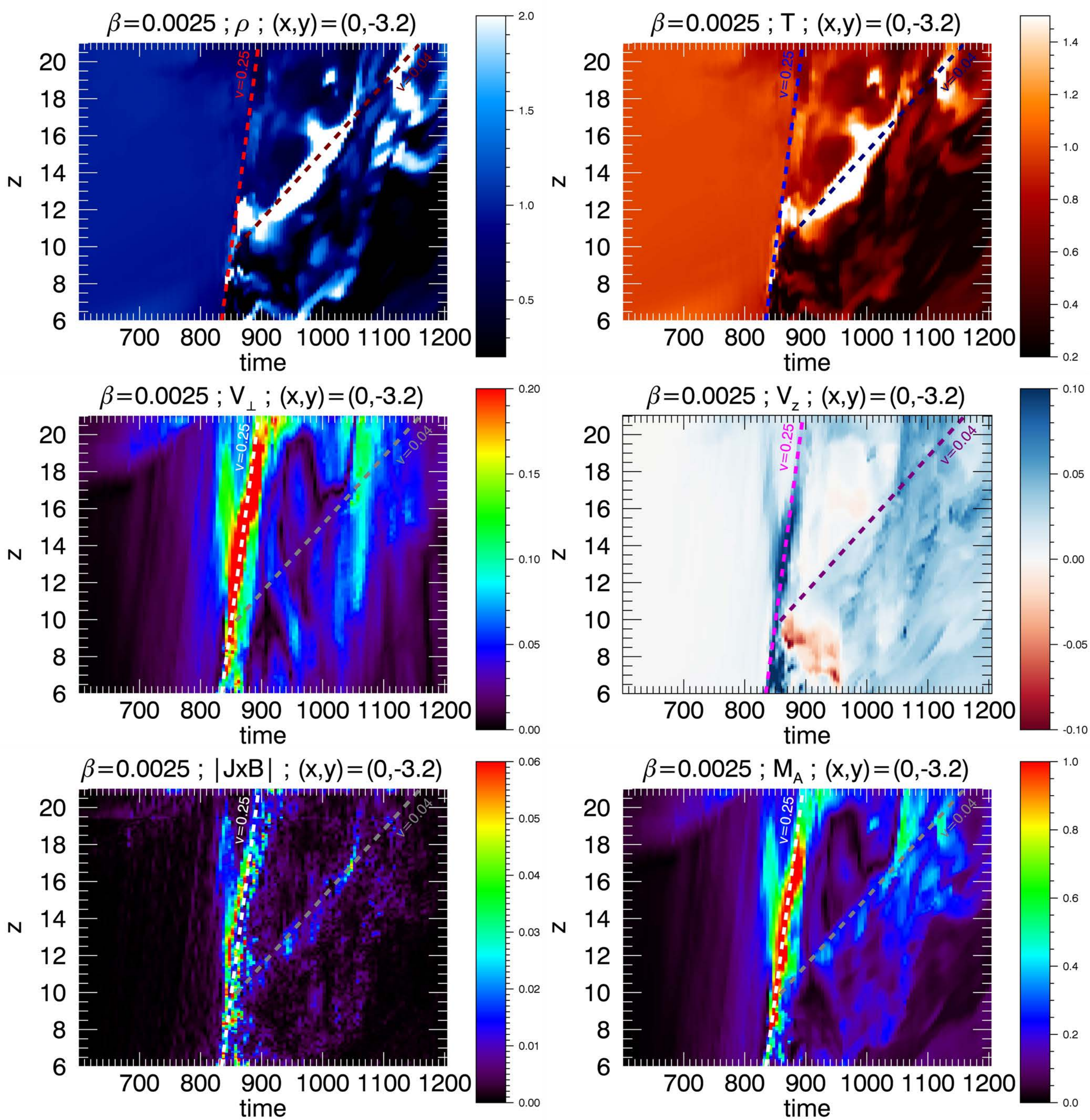}}
\caption{Time evolution of the plasma density, $\rho$, temperature, $T$, vertical and horizontal velocities (respectively $v_z$ and $v_\perp$), magnitude of the Lorentz Force, $|\jj\times\bb|$, and Alfv\'en Mach number, $M_A$, along the vertical direction $z$ at a particular points $(x,y)$, in units of $L_0$, in the $\beta=0.0025$ simulation. The phase velocity $v_\varphi=0.25$ of the propagating wave is indicated by bright dashed lines, while the bulk speed of material transport at $v_z=0.04$ is indicated by darker dashed lines. For comparison, the initial uniform density is $\hat{\rho}=1$, the temperature is $\hat{T}=1$, and the ambient Alfv\'en speed is $\hat{c}_A=0.28$. 
        }
        \label{Fig:TsliceBeta00025} 
\end{figure*}

Figure \ref{Fig:TsliceVelocity} presents the time evolution of the vertical ($v_z$) and horizontal ($v_\perp$) components of the plasma flow along 
the vertical direction at a given point $(x,y)$ in each $\beta$ simulation. The time slice shows the upward propagation of an enhanced velocity 
structure for both components.  In this figure, we note the propagation 
front of the upward-moving wave and compare it to the local plasma flow speeds.  
We recall that the Alfv\'en speed is constant ($c_A=0.28$), while the sound speed, $c_S$, 
decreases with smaller $\beta$ (cf.\ values in Table \ref{Tab:Scaling}). The wave character of the jet is immediately apparent in the 
discrepancy between the phase speed and the bulk speed of the plasma for the three lowest-$\beta$ simulations. The wave propagates upward 
at near-Alfv\'en speed, i.e.\ $v_\varphi \approx [0.26,0.22,0.25]$ respectively for the $\beta=[0.25,0.025,0.0025]$ runs, which represents 
$[93,78,89]\%$ of the large-scale Alfv\'en speed and $[2.0,5.4,19.2]$ times the global sound speed, respectively. The phase speed of the 
propagating wave is Alfv\'enic and strongly supersonic. In these simulations, the actual vertical bulk 
speed of the plasma, $v_z$, shown color-coded in the left column of Fig. \ref{Fig:TsliceVelocity}, 
is significantly smaller than $c_A$, with maximum 
values not exceeding $0.14$ for $z>8$ (higher velocities can be observed elsewhere, in particular around the reconnection site). The horizontal 
speed, $v_\perp$, is roughly the same for the three low-$\beta$ simulations.

The $\beta=1$ simulation differs from the others. In that run, the speed of the upward propagating front is $v_\varphi \approx 0.17$; this is lower than 
in the other cases and, most importantly, corresponds to the vertical bulk speed of the plasma. In this case, the upward motions correspond to the 
bulk flow of the material, and the flow is much slower than the large-scale Alfv\'en and sound speeds (here $c_S=0.26$).

  \begin{figure*}[h!]
\resizebox{0.95\hsize}{!}{\includegraphics[width=\linewidth]{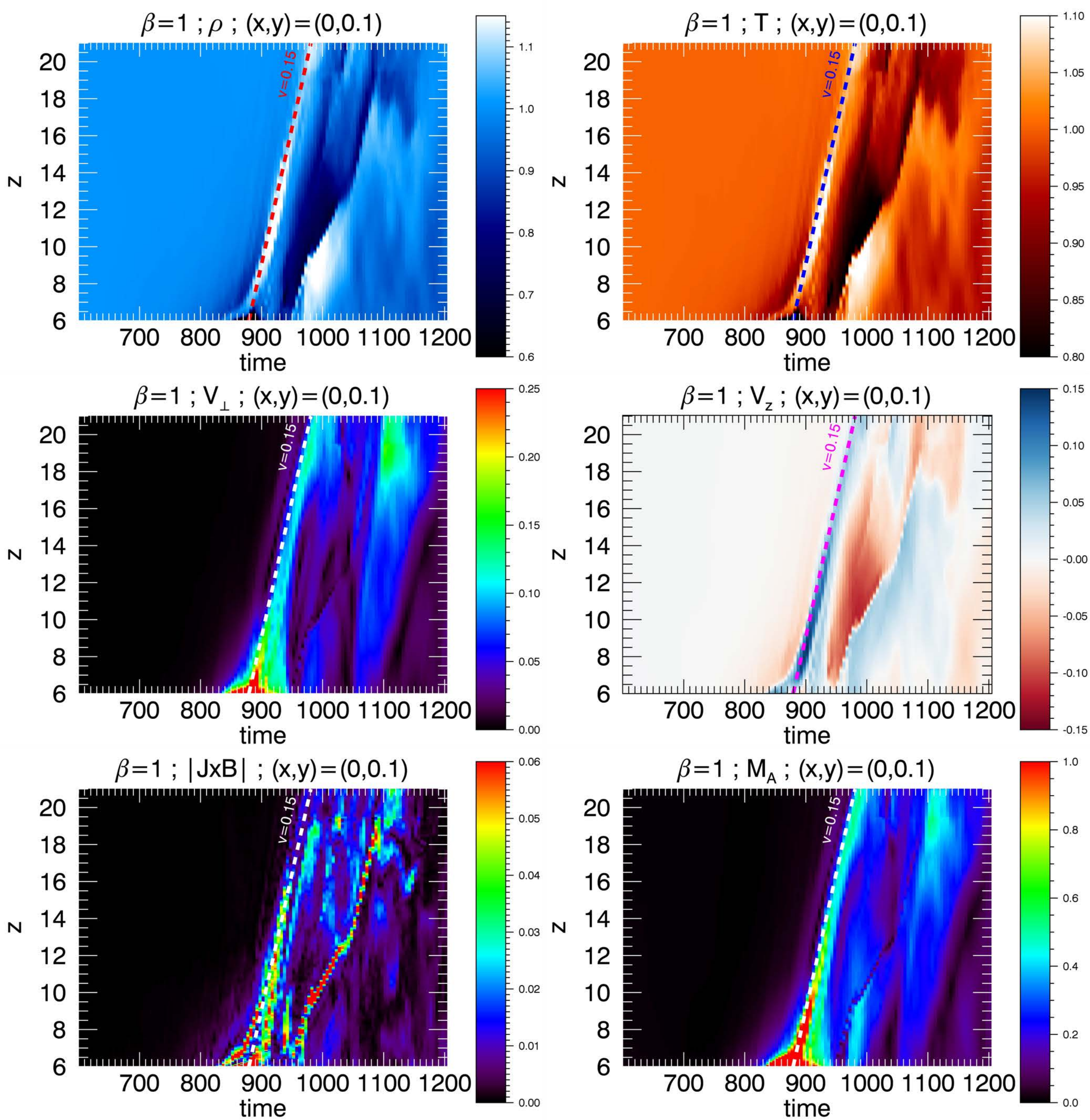}}
\caption{Time evolution of the plasma density, $\rho$, temperature, $T$, vertical and horizontal velocities (respectively $v_z$ and $v_\perp$), magnitude of the Lorentz Force, $|\jj\times\bb|$, and Alfv\'en Mach number, $M_A$, along the vertical direction $z$ at a particular points $(x,y)$, in units of $L_0$, in the $\beta=1$ simulation. The phase velocity $v_\varphi=0.15$ of the propagating wave is indicated by bright dashed lines. For comparison, the initial uniform density is $\hat{\rho}=1$, the temperature is $\hat{T}=1$, and the ambient Alfv\'en speed is $\hat{c}_A=0.28$. 
        }
        \label{Fig:TsliceBeta1} 
\end{figure*}

To examine the differences further, Figs. \ref{Fig:TsliceBeta00025} and \ref{Fig:TsliceBeta1} present time slices of different quantities for the 
$\beta=0.0025$ and $\beta=1$ simulations, respectively. 
Figure  \ref{Fig:TsliceBeta00025} shows the upward displacement of a region of enhanced density and temperature that traces the 
jet in the $\beta=0.0025$ simulation. 
The 1D cut  is located at $(x,y)=(0,-3.2)L_0$ which is roughly at the distance of the photospheric separatrix between the closed and open fields. Relative to Fig. \ref{Fig:BetaDyn}, bottom left panel, this vertical cut is located at roughly the same distance from the centre than the open cyan field lines, and is passing through 
the high density branch of the \hjet which is on the side of the viewer. This branch corresponds to the high density region at $t=875$, $z\sim 12$, in Fig. \ref{Fig:TsliceBeta00025}, top left panel.
The panels display structures propagating at two different speeds. First, a propagation front is present 
with a speed of $0.25$, corresponding to the propagation of the nonlinear torsional Alfv\'en wave. This front spatially corresponds 
to an enhanced perpendicular velocity, to a strong Lorentz force $|\jj \times \bb|$, and to a local Alfv\'en Mach number, $M_A$ (the ratio of the local momentum to the local Alfv\'en speed), close to 1. The surrounding field not having been previously perturbed, the local Alfv\'en speed equals the general large-scale value of $\hat{c}_A=0.28$, at the moment of the passage of the wave. While the wave propagates upward close to the local-Alfv\'en speed, its associated plasma motions attain the local Alfv\'en speed, as shown by the values of $M_A$ close to unity.
The driver of the upward flows is the Lorentz force. The tension of the kinked magnetic field line creates an upward 
and rotating force which accelerates the plasma. Its action may be qualitatively understood by the analytical model of \citet[][cf. Sect. 3 of that study]{Shibata85}. 
The wave propagates upward at a speed several times higher than the bulk flow of the plasma. 
The plasma, which has been accelerated by the passage of the torsional wave, then propagates upward trailing the wave. This explains the second type of structure observed in the time-slice plots of the density and temperature. In the wake of the propagation front, we observe that the high-density and -temperature region moves upward with a speed of $0.04$, which corresponds to the local plasma velocity.   

For the three low-$\beta$ simulations, the jet dynamics result from the action of these two components, the wave propagation and the plasma upflows. Along each reconnected field line, 
the helicity and twist are redistributed and a nonlinear wave is generated. The propagating torsional Alfv\'en wave accelerates, 
heats, and compresses the plasma, giving it a rotating helical shape. Then, the structure evolves in the wake of the wave, due 
to the flow speed imparted to the plasma. The overall 3D morphology of the \ujet is the result of these processes developing at 
multiple points in the domain along the sequentially reconnecting magnetic field lines.

In the $\beta=1$ simulation, while the main driver remains the propagating torsional Alfv\'en wave, the situation is a bit simpler. {The 1D cut  is located at $(x,y)=(0,0.1)L_0$ close to the centre of the domain. Relatively to Fig. \ref{Fig:BetaDyn}, top left panel, this cut is passing in the middle of the open cyan field line and passing through high density region of the \hjet.}
The jet is here formed by a bulk flow co-spatial with the wave. Figure \ref{Fig:TsliceBeta1} shows that the upward displacement of the enhanced-density 
and -temperature region progresses at the local vertical speed of the plasma, here around $0.15$. 
The jet is thus solely the resultant of an upward  
bulk flow. The jet rotates with a transverse speed in the same range as the vertical speed. 
As in the low-$\beta$ runs, the jet is nonetheless 
magnetically driven. This is demonstrated by the fact that the upward-moving material 
propagates at the local Alfv\'en speed, as shown by the value 
of the Alfv\'enic Mach number, $M_A$, which is markedly high at the location 
of the jet, close to unity in the bottom part and slightly decreasing ($>0.6$) as the jet progress further up. Furthermore, the jet is directly cospatial with a region 
of strong Lorentz force, $|\jj \times \bb|$. Hence, the driver of the jet in this case is again the magnetic torsional wave. 
For this $\beta=1$ run, the wave accelerates the bulk of the plasma at the same speed as the phase speed of the wave. 
The helical jet structure, as shown in Fig. \ref{Fig:BetaDyn}, here directly corresponds to and maps the propagating torsional wave.

\section{Morphology of the helical jet} \label{sec:Betamorpho} 

The variations in $\beta$ have another important consequence for the dynamics of the \ujet. 
The morphology of the \hjet is indeed strongly influenced by $\beta$. This is, of course, partly due to the differences in the  
driving properties studied in the previous section. In addition, there are differences in the dynamics at the reconnection site, 
as discussed in this section. 

Beyond the trigger time and energy already defined in Sect. \ref{sec:Trig}, from the 3D data of the time evolution of the thermodynamic quantities, 
we can derive other properties of the jets in the different runs, such as its width, $R$, and its duration, $\Delta t$. To do this, we must first define the 
boundary of the jet within the continuous 3D distribution. For each simulation, we have defined corresponding threshold values of the density, $\rho_t$,
and temperature, $T_t$. These values correspond to regions of steep gradients of the corresponding quantities and define clearly the region of 
increased density and temperature that contains the jet. The values used for each run are listed in Table \ref{Tab:BetaParam}.
The jets presented in Fig. \ref{Fig:BetaDyn} were plotted using the corresponding value of $\rho_t$. 
During the estimation of the different quantities presented hereafter, we checked that they were only minutely affected by reasonable variations of
the precise values of $\rho_t$ and $T_t$.  In particular, different threshold values of $\rho_t$ and $T_t$, to within a factor of 2, 
only marginally changed the estimated quantities. The measurement error bars provided in Table \ref{Tab:BetaParam} are the outcomes 
of these two tests.

The choice of the values of $\rho_t$ and $T_t$ is, however, strongly influenced by the plasma $\beta$. Indeed, the 
pressure, density, and temperature excesses in the jet at lower $\beta$ are relatively larger than in the higher-$\beta$ runs. 
This is a direct consequence of the stronger impact of the magnetic field on the plasma in the lower-$\beta$ environments. 
The \hjet is the result of the compressive effect of the propagating nonlinear Alfv\'enic wave. While the Lorentz force in the kinked part of 
the field lines has a constant magnitude between the different cases, its impact on the plasma dynamics is relatively stronger 
when $\beta$ is smaller, i.e., it induces a stronger pressure increase at lower plasma $\beta$.  
Following an adiabatic ideal gas evolution, the plasma becomes denser and hotter. At lower $\beta$, the system can therefore more efficiently generate a jet that is dense and hot relative to the 
surrounding environment. Therefore, our model predicts that the plasma properties observed in the jet will depend sensitively upon the plasma 
$\beta$ environment and, hence, on the layer of the atmosphere in which the jet is generated (see Sect. \ref{sec:Concl}). It should be emphasized here that this conclusion is only for the helical jet driven by the expolosive burst of interchange reconnection. As discussed above, the early-phase straight jet exhibits the opposite variation with plasma $\beta$.

It is evident from Fig. \ref{Fig:BetaDyn} that while all the jets present a rotating helical structure, the characteristics of the helix vary with $\beta$. 
The lower $\beta$ is, the wider is the jet, i.e., the larger is the amplitude of the helix. For $\beta=1$, the 
spire of the jet is compact and the jet appears as a thin and very collimated structure. In the case of 
the low-$\beta$ runs, the jets present a much wider structure similar to a rotating hollow cylinder or a {\it 
"cylinder with helical structure on the surface"} as described in \citet{ShenY11}.
The half-width of the jet, i.e.\ the amplitude of unwinding helical global wave, is taken as the radius of the 
hollow cylinder, $R$, given in Table \ref{Tab:BetaParam}. The ratio of $R$ to the 
characteristic size of the closed domain (the radius of the 
fan separatrix at the bottom boundary, equal to $2.2$) is respectively on the order of
$[1.1, 1.2, 1.4, 1.4]$ from higher to lower $\beta$. While the $\beta=1$ \hjet appears narrower, 
it is nonetheless moving transversely across the domain. The whole structure is dynamically displaced 
horizontally over a distance comparable to the scale of the closed domain. This is likely due to the fact that the $\beta=1$ jet is more directly 
advected by, or embedded with, the propagating wave: as the reconnection site moves sideways, so does the jet structure.

We also estimate the duration of the jet, $\Delta t/t_0$, by inspecting the period during which a jet structure (with $\rho \ge \rho_t$ and $T \ge T_t$  
is present in the simulated data (e.g., as in PAD09, cf. their Figs. 4 \& 5). 
The obtained values are listed in Table \ref{Tab:BetaParam}. As was done for $R$, the error bars 
are derived from the results using the different threshold values of density and temperature. 
The variations in jet duration between the different runs can also be estimated from the energy plots presented in Fig. \ref{Fig:BetaE}. 
Overall, one notes that the jets tend to last longer for lower values of the plasma $\beta$, with the $\beta=1$ jet being markedly briefer 
than the three other cases.

In Table \ref{Tab:BetaParam}, we also list the average velocities measured in the jet for the different runs.
The vertical velocity, $\langle v_z \rangle$, roughly corresponds to the axial velocity along the direction of the jet 
(accounting for the inclination angle of $10^\circ$ modifies these values only very slightly). The
characteristic transverse velocity, $\langle v_{\perp} \rangle$, is taken to be the characteristic velocity in the $xy$ plane 
in the middle of the jet (i.e., for $z > 8$). To determine these average values, 
we took the mean of the values of the velocity, restricted to the 3D sub-volume of high density and high temperature that defines the 
jet (i.e., using the threshold values in density, $\rho_t$, and temperature, $T_t$, defined above). We checked that varying 
the density, the temperature, or the combination of the two does not change significantly the derived values of the average velocities. 
As with $R$ and $\Delta t$, the measurement error bars provided for the average values are the outcome of the different tests varying 
$\rho_t$, and $T_t$.

The total velocity of the plasma, 
$\langle v_{plas} \rangle$, in Table \ref{Tab:BetaParam} is the quadratic sum of the two velocities and corresponds to the bulk flow of the plasma.
The jet phase velocity, $v_{\varphi}$, is taken as the vertical speed of the propagation front 
of the nonlinear Alfv\'enic wave, i.e., its phase speed, as estimated in Sect. \ref{sec:Betadriver}. 

We note that the (non-dimensional) perpendicular average velocity, $\langle v_\perp \rangle$ in the jet is relatively constant for all the 
runs, equal to $\approx 0.08 $, while $\langle v_z \rangle$ decreases by a factor of 2 from the highest-$\beta$ to the lowest-$\beta$ simulation.   
These average values correspond to the mean velocities of the plasma within the whole jet structure, and thus are significantly smaller than 
the maximum values observed in Fig. \ref{Fig:TsliceVelocity}. For the high-$\beta$ run, the average velocities are higher as the jet 
is co-spatial with the Alfv\'enic torsional wave.
For the low-$\beta$ runs, in contrast, the average values are dominated by the velocities of the plasma after it has been accelerated by the nonlinear wave. 
In any case, the phase velocity of the jet, $v_\varphi$, is much larger than the plasma velocity in all the runs. Apart from the $\beta=1$ run, 
$v_\varphi$ is above $0.8 c_A$. Both $v_\varphi$ and $\langle v_{plas} \rangle$ appear to depend only very weakly upon $c_S$, especially in the $\beta = 0.25$ to $0.0025$ range. 

From the estimated average transverse velocity, $\langle v_\perp \rangle$, radius, $R$, and jet duration, $\Delta t$, we can derive the non-dimensional 
angular velocity of the rotation within the jet, $\omega t_0$, the pitch of the helical structure, $h/L_0$, and an estimate for the numbers of turns 
in the jet, $\Delta N$, from its morphological properties: 
\begin{align}
\omega t_0& = \f{\langle v_\perp \rangle}{R}, \\
\f{h}{L_0} & = 2\pi\f{ R \langle v_z \rangle}{\langle v_\perp \rangle}=2\pi\f{ \langle v_z \rangle}{ \omega}, \\
\Delta N & = \Delta t \f{\omega}{2\pi} = \Delta t \f{\langle v_z \rangle}{h}. 
\end{align} 
These estimations of the properties of the jet are regularly done in observed cases \citep[e.g.,]{HongJC13}. 
We treat our data in a similar way, which allows a direct comparison between the properties of the present simulated jets with 
observed jets, as discussed in Sect. \ref{sec:Concl}.

\begin{table*}[h!]
\caption{Characteristics of the helical jets in the parametric $\beta$ simulations:  Sound speed, $c_S$; 
free magnetic energy, $E_{trig}$, at the trigger time, $T_{trig}$; 
jet duration, $\Delta t$; radius, $R$; threshold density, $\rho_t/\rho_0$, and temperature, $T_t/T_0$, ratios used to define the jet relative to 
initial values in the open field; average vertical, $\langle v_z \rangle /V_0$, transverse, $\langle v_\perp \rangle /V_0$, and total, $\langle v_{plas} \rangle /V_0$, velocities of the 
bulk flow measured in the jet; phase 
speed of the jet, $v_{\varphi}/V_0$; derived non-dimensional angular velocity, $\omega\ t_0$, helical pitch, $h/L_0$, and ejected 
number of turns of twist, $\Delta N$.}
\label{Tab:BetaParam} 
\centering
\begin{tabular}{ccccc}
\hline 
\hline 
$\beta$ & $1$ & $0.25$ &  $0.025$ & $0.0025$ \\
$c_S$  &  $0.26$    &  $0.13$      & $0.041$   &  $0.013$ \\
\hline 
 $E_{trig}/E_0$ &  $43.0 \pm 0.05$     &  $35.3 \pm 0.05$     & $32.1 \pm 0.05$     &   $30.4 \pm 0.05$     \\
$T_{trig}/t_0$ &  $860 \pm 5$     &  $795 \pm 5$     & $765 \pm 5$     &  $755 \pm 5$      \\
\hline 
$\rho_t/\rho_0$&  $\sim 1.1$      &  $\sim 1.2$       & $\sim 1.6$       &  $\sim 2$         \\
$T_t/T_0$     &  $\sim 1.06$      &  $\sim 1.2$      & $\sim 1.3$       &  $\sim 1.5$        \\
$R/L_0$         &  $2.5 \pm 0.3$  &  $2.7 \pm 0.3$   & $3.0 \pm 0.3$   &  $3.1 \pm 0.3$      \\
$\Delta t/t_0$  &  $175 \pm 50$     &  $250 \pm 50$     & $325 \pm 50$     &  $325 \pm 50$      \\
\hline 
$\langle v_z \rangle /V_0$       &  $0.07 \pm 0.02$  &  $0.07 \pm 0.02$  & $0.04 \pm 0.02$  &  $0.03 \pm 0.02$   \\
$\langle v_\perp \rangle /V_0$   &  $0.08 \pm 0.02$  &  $0.09 \pm 0.02$  & $0.07 \pm 0.02$  &  $0.07 \pm 0.02$   \\
$\langle v_{plas} \rangle /V_0$  &  $0.11 \pm 0.03$  &  $0.11 \pm 0.03$  & $0.08\pm 0.03$  &  $0.08 \pm 0.03$   \\
$v_\varphi/V_0$ &    $0.17 \pm 0.03$  &  $0.22 \pm 0.03$  & $0.25\pm 0.03$  &  $0.25 \pm 0.03$ \\
\hline 
$\omega t_0$    &  $0.032 \pm 0.012$&  $0.033 \pm 0.011$& $0.023 \pm 0.009$&  $0.022 \pm 0.009$ \\
$h/L_0$         &  $14 \pm 9$      &  $13 \pm 8$      & $11 \pm 9$      &  $8 \pm 9$       \\
$\Delta N$  &  $0.9 \pm 0.6$    &  $1.3 \pm 0.7$    & $1.2 \pm 0.6$    &  $1.2 \pm 0.7$      \\
\hline 
\end{tabular}
\end{table*}

We observe that the non-dimensional angular speed $\omega t_0$ tends to decrease with decreasing $\beta$. Since the transverse average jet speed, 
$\langle v_\perp \rangle$, tends to be constant, this is mostly an effect of the larger width of the jet. We note that the non-dimensional pitch angle, 
$h/L_0$, also tends to decrease with decreasing $\beta$ (however, observe the large uncertainty in 
these values). As noted earlier, this pitch-angle change is also verified by a visual inspection of the morphological shape of the jet 
(see Fig. \ref{Fig:BetaDyn}). The higher the $\beta$, the thinner and less helical the jet appears.
 From the output of the simulations, we note that the density structure is rotated by half a turn along a 
height of $14$ units for the $\beta=1$ jet and of $8$ units in the $\beta=0.0025$ case. The variation in the pitch angle is 
mainly determined by the ratio $\langle v_z \rangle / \langle v_\perp \rangle$. Since the higher-$\beta$ simulations are relatively more efficient at accelerating 
the plasma upward, compared to the transverse acceleration, these jets appear more pitched.

Excluding the $\beta=1$ run, we note that
 the ejected twist $\Delta N$ inferred from the rotation is roughly constant for the runs at 
lower $\beta$ and equal to $\approx 1.2$. This value for $\Delta N$ is fully consistent with the amount of twist/helicity 
injected in the system by the boundary motions, which is on the order of $1.2$ turns (cf.\ Sect. 3 in PDD15). As in 
PAD10, we note that here also the system appears quasi-potential after the jet occurs, i.e., only 
low-lying field lines next to the inversion line in the closed domain remain twisted. As noted in PAD09,  
for the lower-$\beta$ runs, the reconnection at the 3D null during the jet removes 
most of the helicity by transferring it to the open field. In \citet{Pariat15b}, we measured that 
$90$\% of the helicity was eventually removed from the closed system. The amount of twist that we 
derive here from the observed rotation in the \hjet is fully consistent with this picture. 
We emphasize that the derivation of the twist from the rotation of the jet is completely independent of the magnetic field 
measurements. It indicates that, in observational cases, the derivation of azimuthal velocities 
assuming a cylindrical structure can be used to estimate the amount of twist stored initially.

It is interesting to note that, at any given time, the \hjet never displays a full rotation (e.g., Fig.\ref{Fig:BetaDyn} and other figures 
in previous studies). It is only the time-integrated observation of the jet evolution that allows one to infer the stored twist. 
This can be qualitatively understood since only about half of the twist contained on the closed field 
lines is eventually transferred to the open field lines when they reconnect. Consider a closed field line with a 
total twist $\tau$: if this field line reconnects with the open field in its middle, only $\tau/2$ will 
be acquired by the newly formed open field line and ejected. The other half of the twist will 
remain behind on a newly formed closed field line. Such field lines can later reconnect and transmit, for example, twist $\tau/4$ 
to the open field. Hence, at any given time, only a fraction of the total stored helicity is given to the reconnected field lines.
It is only due to the continuous and sequential reconnection that the newly open field lines extract nearly all the 
helicity from the closed field. The ability of the reconnection site to move within the 3D volume and reconnect most of the 
twisted flux enables the most efficient release of the helicity and energy. In the lower-$\beta$ runs, we indeed observe that the 
reconnection site moves dynamically in the 3D space, hence more field lines sequentially reconnect, in some cases 
multiple times.  The jets have thus a longer duration and their final free energy is notably lower, as can be noted from Fig. \ref{Fig:BetaE},
compared to the $\beta=1$ case.

For the $\beta=1$ run, the amount of twist derived from the rotation 
is significantly smaller ($\Delta N \approx 0.9$), even though the \hjet was triggered later than 
in the other runs and, hence, after a larger amount of helicity has been stored ($\approx$ 0.1-0.2 turns more).
Thus, the $\beta=1$ system releases much less helicity and energy than the 
other cases, as indicated by the larger amount of free energy remaining in the system (cf.\ Fig. 
\ref{Fig:BetaE}) compared to the other runs. It can also be noted visually (not shown here) 
that many more twisted field lines are still present in the $\beta=1$ system at the end of the \hjet phase. 
Here again, the measurements of the rotation speed in the jet allow a quantitative estimate of the amount of helicity ejected. 
The smaller value of $\Delta N$ indicates that, for $\beta=1$, the system is much less 
efficient at releasing helicity compared to the other runs. We observe that the reconnection site 
moves less in the 3D space compared to the lower-$\beta$ simulations. 
This is probably a consequence of the plasma resisting more strongly the rotation of the reconnection site, for $\beta=1$. 
Because of the stronger plasma pressure, the 3D null point is not able to move as easily in the domain, and the 
reconnection of the twisted field lines over a large volume is inhibited. The reconnection site accesses less twisted flux, 
less flux can reconnect, and thus less helicity and free energy are removed from the closed system.

The stronger impact that the magnetic field has at lower $\beta$ is thus the likely reason 
for the different morphology of the jet. The shape of the \hjet is the combined consequence 
of, first, the sequential interchange reconnection of the twisted field lines and, second, 
the ability of the propagating nonlinear Alfv\'enic wave to 
compress and accelerate the plasma (as discussed in Sect. \ref{sec:Betadriver}). The helical 
structure is induced by the displacement of the reconnection site in the 3D volume. However, 
the extension of the current sheet, its rotation, and the displacement of the reconnection 
site are less easily achieved as the plasma $\beta$ increases and the plasma is better able to resist 
the magnetic forces. Comparing the dynamics of the current sheet in the 
different simulations indeed shows that the current sheet is less extended and rotates 
over a smaller portion of the 3D domain for higher values of $\beta$. Consequently, 
the sequentially reconnected field lines remain in a smaller volume for higher 
$\beta$ and the jet appears more compact. In contrast, for lower $\beta$ the reconnection 
site is displaced over a very large domain. This leads to the reconnection of field lines farther 
away from the centre of the configuration and creates a jet with the 
shape of a helical hollow cylinder or curtain, as observed in Fig. \ref{Fig:BetaDyn}.

\section{Conclusion} \label{sec:Concl} 

The present study has further explored the physics of a model for the generation of solar jets \citep[][PAD09]{Pariat09a}. 
We have presented a parametric study of the influence of the plasma $\beta$ parameter that extends our previous 
parametric studies of inclination angle and photospheric flux distribution \citep[][PDD15]{Pariat15a}.  
We have confirmed that the model of PAD09 is robust, and can  lead to the generation of both 
\gjets and \hjets, for a wide variety of uniform atmospheric conditions with $\beta$ ranging from $10^{-3}$ to $1$.  We observed 
that the dynamics of the jets at $\beta=0.025$ are very similar to those at $\beta=0.0025$. We thus expect that this model is also 
working for values lower than $10^{-3}$, hence for any values of $\beta \le 1$. The model presented here is thus potentially able to explain jet-like events occurring 
in all the different layers of the solar atmosphere where the standard conditions for the validity of MHD are met. 

While our model, based on a fully 
ionised single-fluid plasma with no atmospheric stratification, can properly represent large-scale coronal events, by no means does it 
completely represent the physics of jets that develop in the lower layers of the solar atmosphere.
The fact that our jet model is able to produce \hjets for a wide range of $\beta$ is  important, given that helical structures 
are observed at many different scales in the various layers of the solar atmosphere. As discussed in the Introduction, twist and other signatures of rotating motions 
have been ubiquitously observed in jet-like events from the photosphere to the corona 
\citep[e.g.,][]{Patsourakos08,Curdt10,DePontieu12,Tian14,Cheung15}.  Our study suggests that 
a universal mechanism could potentially explain the helical properties observed in all 
types of solar jet-like phenomena.  This mechanism, the generation of \ujets induced by sequential reconnection and 
driven by propagating torsional Alfv\'enic waves, appears to fit multiple observed properties of the large-scale coronal jets, and could likely contribute
to the similar dynamics observed in spicules and chromospheric jets. This idea, however, must be tested and evaluated 
using more complete numerical models of the photospheric and 
chromospheric layers \citep[e.g.,][]{MartinezSykora09,MartinezSykora11,MartinezSykora13,Kitiashvili13,Takasao13,YangL13b}.

We have observed that varying the plasma $\beta$ modifies the efficiency of 
the forcing of the driving motion on the 3D null point. Smaller $\beta$ values 
lead to more efficient forcing of the 3D null, since the greater volume expansion of the closed domain induces earlier and stronger reconnection. However, at low $\beta$ this does not 
necessarily induce a high-density outflow, and the \gjet is actually weaker at lower 
values of $\beta$. Although the reconnection is stronger at lower $\beta$, the \gjet is more marked  
at higher $\beta$ (Sect. \ref{sec:Trig}). 

Our results concerning the trigger of \gjets and its influence on the generation of \hjets (Sect. \ref{sec:Trig}) thus confirm and supplement the main conclusion drawn in PDD15 (summarised in Sect. \ref{sec:PaperI} of the present study). Whether or not the \gjet is actually observed during the pre-helical-jet phase, at lower plasma $\beta$ we note a higher amount of reconnection during the pre-helical-jet phase associated with an earlier trigger of the \hjet. Similar to the results obtained from the two previous parametric simulations of PDD15, we note that the occurrence of strong reconnection during the pre-helical-jet phase has an important impact on the trigger of the \hjet. Thanks to the present parametric simulation, we note that the presence of a high-density outflow (the \gjet) is not the determinant factor; rather it is the occurrence of reconnection prior to the \hjet.
The present parametric study confirms  that the stronger is the reconnection during the 
\gjet phase, the lower is the energy/helicity threshold for triggering the instability of the \hjet, i.e., the easier it is to generate the \ujets.
As discussed in the conclusion of PDD15, this result is very revealing for the instability leading to the trigger of \hjets.

The parametric study carried out here, however, provides original results that go beyond the results of PDD15 and previous studies. It highlights for the first time the direct impact of the plasma $\beta$ on the morphology and plasma properties of the \hjet (Sect. \ref{sec:Betamorpho}).
 At lower $\beta$, the jet assumes the shape of a high-density helix on the surface of a hollow cylinder. For $\beta \approx 1$, the jet is much more compact and much more collimated, but nonetheless exhibits a dynamic transverse displacement in the domain. Its ``Eiffel Tower'' shape is 
more distinct. Higher $\beta$ induces a smaller width of the global cylindrical volume and a shallower helical pitch angle. 
Jet at higher $\beta$ thus appear more collimated than at lower $\beta$. 
Low-$\beta$ helical jets also have a higher relative density and temperature compared to their environment. 

Apart from the $\beta=1$ case, the amount of twist ejected is only weakly influenced 
by the plasma $\beta$. The 3D null point is always 
able to efficiently eject a substantial amount of magnetic helicity, on the order of one turn 
of the magnetic field. Several observational studies have investigated in detail the kinematics of 
the jet plasma (Sect. \ref{sec:Introduction}).  Assuming a cylindrical rotation, the jet radius, $R$, the transverse velocity, $v_t$, 
the rotation rate, $\omega$, and the number of turns, $\Delta N$, have been derived in observational cases, in a way similar to 
our derivations of these quantities in Sect. \ref{sec:Betamorpho}. Interestingly, the published results are very consistent 
independent of the scale. From the amplitude of the motion and the transverse 
velocity observed in a chromospheric jet \citep[][their Fig.\~4]{LiuW09}, a rotation rate of 
$\omega \approx [0.011-0.025] \U{rad\ s^{-1}}$ of the untwisting structure was deduced. 
For a large coronal \bjet,  \citet{ShenY11} found a 
rotation rate $\omega \approx 0.011 \U{rad\ s^{-1}}$ and an ejected twist $\Delta N \approx [1.2-2.5]$ turns. 
\citet{ChenHD12} estimated $\omega \approx [0.01-0.015] \U{rad\ s^{-1}}$ while \citet{HongJC13} obtained 
$\omega \approx 0.014  \U{rad\ s^{-1}}$ and $\Delta N \approx 0.9$ turns. 
\citet[][cf. Fig. 5]{Cheung15} analysed a transition region jet with a width of about $10 \U{Mm}$ and transverse velocities 
above $50 \U{km\ s^{-1}}$, hence a rotation rate larger than $0.01 \U{rad\ s^{-1}}$.  Using the scaling of $t_0$ 
suggested by our Table \ref{Tab:Scaling}, the equivalent dimensional rotation rate $\omega$ measured in our simulation 
ranges between $0.01$ and $0.03 \U{rad\ s^{-1}}$, with an ejected number of 
turns $\Delta N \approx [0.9-1.3]$.  The rotation inferred from our modeled \hjet, therefore, is very consistent with the 
properties of these observed jets. 

The $\beta=1$ case, which corresponds to the lower layers of 
the solar atmosphere, is significantly less efficient at removing magnetic helicity. Less twist 
is ejected and larger amounts of helicity and energy remain in the 
system following the jet. Previously, we showed that the 3D null-point configuration at the base of the present jet model readily 
allows the generation of recurrent homologous events \citep[][PAD10]{Pariat10}. When subjected to a constant energy input, the system produces quasi-periodic jets. Here 3D null points can play the role of a magnetic ``capacitor'' and efficiently
store free magnetic energy in the closed domain. Each jet corresponds to a phase of free-energy release, during which the system 
relaxed partially toward its minimum-energy state. When subjected to a constant energy input, the 
system is able to produce quasi-periodic jets. For the $\beta=1$ runs, more energy/helicity/twist is left in the system; therefore 
the system remains closer to the instability threshold. If energy were 
continuously injected into this system, as in PAD10, we conjecture that jets would be generated 
at a much higher frequency for $\beta=1$ than for lower $\beta$. This may explain the 
much higher occurrence rate of jet-like phenomena in higher-$\beta$ environments. Assuming that $\beta=1$ can, to some extent, model the 
generation of solar spicules, our results could explain the very high occurrence rate and recurrence of spicules compared to 
larger-scale jet-like events. 

In the present study we have also introduced a new analysis of the physical mechanism
driving the plasma in our simulation. The use of time-space diagrams enabled us to obtain a clearer understanding of the underlying driver of the \ujets (Sect. \ref{sec:Betadriver}). We confirmed previous results \citep{Shibata85,Shibata86,Canfield96,Jibben04,Torok09,Pariat09a,Lee15} 
that the primary driver of the \hjets is the propagating nonlinear Alfv\'enic 
torsional wave that develops on the sequentially reconnected open field lines. 
Compared to previous studies, our parametric study revealed how the plasma $\beta$ of the surrounding field could influence the properties of the \ujet.
We found that, at all $\beta$, the acceleration was due to 
the Lorentz force present in the kinked section of the newly reconnected open field lines. The Lorentz force induced a local acceleration 
of the plasma at a velocity close to the local Alfv\'en speed. As the field lines untwist, the propagation of the twist induces a vertical as well as an 
azimuthal motion of the plasma. In addition, the wave generates an adiabatic heating and compression of the plasma, with higher 
efficiency for lower values of $\beta$. 
 
At $\beta=1$, we noted that the accelerated plasma and the wave are embedded within each other, and the jet simply corresponds to a 
bulk flow of plasma.  At lower plasma $\beta$, we observed that the propagation/phase speed of the wave was close to the 
ambient Alfv\'en speed, and was much higher that the bulk flow speed of the plasma. Once accelerated at the front of the propagating wave, 
the dense and hot plasma then moves independently of the wave at its own speed along the field lines. 
At low $\beta$, the overall morphology of the jet is thus distinct from the evolution of the kink present on the field lines. 

As discussed in the Introduction, jets are likely generated by multiple acceleration mechanisms, including both the \ejets and \ujets.  The coexistence of the evaporation and \ujets implies multi-velocity observations in jet events. This can possibly explain the discrepancy between the average velocity measurements of coronal 
jets obtained from imaging instruments compared with those obtained from spectroscopy. 
Spectroscopic measurements, which allow estimates of the bulk flow of the plasma in jets \citep{Kamio07,Kamio10,Madjarska11,Young14a,Young14b}, 
rarely found velocities higher than $300 \U{km\ s^{-1}}$. However, for the low-$\beta$ cases, we stress that the untwisting mechanism, by itself,  produces two types of velocities: 
a phase speed that we find to be close to the Alfv\'en velocity of the open field, and a bulk plasma flow 
that is only fraction of the phase speed. Velocity measurements obtained from imaging instruments, based on 
the estimation of a structure-front speed and, hence, more likely to measure the phase speed of a wave, frequently measure 
velocities higher than $500 \U{km\ s^{-1}}$ \citep{Shimojo96,Cirtain07,Savcheva07}. The higher speeds measured by imaging techniques may 
simply reflect the higher phase speed of the \hjet wave front, compared to the much slower bulk plasma flows. The latter may be formed from either the \ejets or the bulk-flow component driven by the untwisting mechanism, or both.

 In summary, the results of our present analysis allows further understanding of the dynamics of jets developed in previous studies. The jet structure and dynamics are the result of the following processes. 
First, the 3D sequential reconnection of the closed, 
twisted field lines creates new open field lines with a large amount of twist close to the footpoint. For each of these new open field lines, 
the twist is then ejected through the generation of propagating torsional Alfv\'enic waves. The propagating wave heats and compresses 
the plasma as it propagates upward at its Alfv\'enic phase speed. The accelerated plasma can then eventually move along the field lines at 
the speed it acquired. While tension-driven motions are observed to be embedded within the \hjet structure, they play only a minor role in 
explaining the dynamics of the plasma. We reiterate that we do not treat any effects related to evaporation jets since our model 
does not include non-ideal plasma effects such as heat conduction or plasma pressure differences between the 
closed and open field, all of which can drive additional flows. A real jet is likely the combination of these different types of 
mechanisms that induce multi-thermal and multi-velocity features. 

Thanks to the recent numerical studies of jets, understanding of the possible underlying jet mechanism has grown quickly. 
Using multi-wavelength EUV spectroscopic observations, \citet{Matsui12} provided the interesting results that most of the velocities measured 
at higher temperature $(T>10^{5.5} \U{K})$ were consistent with \ejets. On the other hand, emissions at lower temperatures were 
much higher than what is expected due to the evaporation mechanism.  Since emission at these lower temperature tends to more frequently 
and more clearly display helical motions, this would suggest that it is mainly in this temperature range that the \ujets mechanism 
that we are modeling here is dominant. In any case, the acceleration mechanism can only be understood fully if a complete 
diagnostic of the jet and surrounding plasma properties is performed. Spectroscopic measurements, such as those provided by the 
IRIS spacecraft \citep{DePontieu14b}, are the keys to advance our understanding of jet-like events in the different layers of the solar atmosphere.

\begin{acknowledgements} 
The authors thank the referee for helpful comments which improved the clarity of the paper.
The authors acknowledge access to the substantial HPC resources of CINES under the allocations 2014--046331, 2015--046331, and 2016--046331 made by GENCI (Grand Equipement National de Calcul Intensif). We also appreciate the support of the International Space Science Institute and the contributions by other team members during the workshops {\it Understanding Solar Jets and their Role in Atmospheric Structure and Dynamics}. K.D.\ acknowledges support from the Computational and Information Systems Laboratory and the High Altitude Observatory of the National Center for Atmospheric Research, which is sponsored by the National Science Foundation. C.R.D., S.K.A., and J.T.K.\ all gratefully acknowledge support from NASA's LWS TR\&T and H-SR programs.
\end{acknowledgements}

\bibliographystyle{aa}  
\bibliography{Jets_param}       
\IfFileExists{\jobname.bbl}{}  
{ 
\typeout{} 
\typeout{****************************************************} 
\typeout{****************************************************} 
\typeout{** Please run "bibtex \jobname" to obtain}  
\typeout{**the bibliography and then re-run LaTeX}  
\typeout{** twice to fix the references!} 
\typeout{****************************************************} 
\typeout{****************************************************} 
\typeout{} 
 }

\appendix

\section{Scaling of the MHD simulations relevant for solar-atmosphere like conditions} \label{An:Scaling}

For a non-dimensional quantity, $\hat{f}$, derived by solving the numerical MHD equations, it is straightforward to derive the corresponding 
dimensional units $f_0$ such that $f_0=\hat{f} \times f_S$, with $f_S$ a characteristic scale of that quantity. 
The scaling of the system is fully determined once the characteristic dimensional density $\rho_0$, pressure $P_0$, temperature $T_0$, magnetic field $B_0$, and length $L_0$ are given. Assuming 
a given value of the plasma $\beta$ and that the gas follows the ideal gas law, only two among $\rho_0$, $P_0$, $T_0$, and $B_0$ need to be specified.
Indeed, these quantities are linked such that:
\begin{gather}
\beta = \f{2\mu_0P_0}{B_0^2}=\f{2\hat{\mu}\hat{P}}{\hat{B}^2} \Leftrightarrow  1 = \f{\mu_{S}P_S}{B^2_S}\\
R_0= \f{P_0}{\rho_0 T_0}
\end{gather}
with $\mu_0=4\pi\times 10^{-7} \U{Wb\ A^{-1}m^{-1}}$ the vacuum permeability  and $R_0$ the ideal gas constant scale.  For a fully ionized plasma composed only of hydrogen, one has $R_0=1.650\times10^4 \U{m^2\ s^{-2}\ K^{-1}}$. 
Fixing two of the quantities $\rho_0$, $P_0$, $T_0$, and $B_0$, therefore, determines the other two. This also fixes the velocity scale, $V_0$, since 
\begin{align}
V_0&=\f{c_S}{\hat{c}_S}=\sqrt{\f{R_0 T_0}{\hat{R}\ \hat{T}}} \\
 &=\sqrt{R_S T_S}=\sqrt{\f{P_S}{\rho_S}}=\f{B_S}{\sqrt{\mu_{S} \rho_S}} \\
 &=\f{c_A}{\hat{c}_A}=\f{B_0\sqrt{\hat{\mu} \hat{\rho}}}{\hat{B}\sqrt{\mu_{0} \rho_0} }.
\end{align} 
In order to close the system either the length scale, $L_0$, or the time scale, $t_0$, must be specified, the two being related such that $L_0=V_0 t_0$.

In our numerical simulation, we use $\hat{B}=\hat{\rho}=\hat{T}=1$, $\hat{R}=0.01$, $\hat{\mu}=4\pi$ and $\hat{P} \in [4\times10^{-2},10^{-2},10^{-3},10^{-4}]$. The choice of $\hat{P}$ determines the value of $\beta$. Then, fixing two of the quantities $\rho_0$, $P_0$, $T_0$, and $B_0$, as well as choosing either the length $L_0$ or the time scale $t_0$, fully determines all dimensions of the MHD system. The following tables \ref{Tab:BetascalingT}, \ref{Tab:BetascalingB}, \ref{Tab:BetascalingB2}, and \ref{Tab:BetascalingL} illustrate a wide range of possible systems that our simulations can represent.  Table \ref{Tab:BetascalingT} shows the effect of using different temperature $T_0$; Tables \ref{Tab:BetascalingB} and \ref{Tab:BetascalingB2} show the effect of using different volume magnetic field $B_0$, fixing either the time scale or the length scale, respectively; and \ref{Tab:BetascalingL} shows the effect of using different length scales, fixing $P_0$ and $T_0$ for each $\beta$. 

The quantities $\rho_0$, $P_0$, $T_0$, $B_0$ correspond to the values in the volume, away from the central polarity. The maximum field in the center of the polarity is equal to $\simeq 14B_0$. The flux in the central polarity is $\simeq 30 B_0L_0^2$. The null is at a height of $\simeq 2.2L_0$, the polarity inversion line at $\simeq 1.6L_0$, and the closed-domain separatrix at $\simeq 3.4L_0$ from the center.

\begin{table*}[h!]
\caption{Possible scaling for the different plasma $\beta$ simulations using constant volume magnetic field $B_0$ (in G) and time $t_0$ (in s), while varying the temperature $T_0$ (in K). The other quantities are: length $L_0$ (in Mm), pressure $P_0$ (in Pa), density $\rho_0$ (in$\U{kg\ m^{-3}}$), velocity $V_0$ (in$\U{km\ s^{-1}}$), Alfv\'en speed $c_A$ (in$\U{km\ s^{-1}}$), sound speed $c_S$ (in$\U{km\ s^{-1}}$), and energy $E_0$ (in J).
}
\label{Tab:BetascalingT} 
\centering
\begin{tabular}{ccccccccccc}
\hline 
$\beta$ & $L_0 $ & $t_0$ & $B_0$ & $P_0$ & $\rho_0$ & $T_0$ & $V_0$ & $c_A$ & $c_S$ & $E_0$ \\
\hline
$ 1.0 $ & $ 0.045 $ & $ 1 $ & $ 3.5 $ & $ 0.049 $ & $ 5.9 \times 10^{-10} $ & $ 10^{4} $ & $ 45 $ & $ 12 $ & $ 11 $ & $ 1.1 \times 10^{14} $ \\ 
$ 1.0 $ & $ 0.1 $ & $ 1 $ & $ 3.5 $ & $ 0.049 $ & $ 5.9 \times 10^{-11} $ & $ 10^{5} $ & $ 143 $ & $ 40 $ & $ 37 $ & $ 3.6 \times 10^{15} $ \\ 
$ 1.0 $ & $ 0.5 $ & $ 1 $ & $ 3.5 $ & $ 0.049 $ & $ 5.9 \times 10^{-12} $ & $ 10^{6} $ & $ 454 $ & $ 128 $ & $ 117 $ & $ 1.1 \times 10^{17} $ \\ 
$ 1.0 $ & $ 0.6 $ & $ 1 $ & $ 3.5 $ & $ 0.049 $ & $ 3.0 \times 10^{-12} $ & $ 2.0 \times 10^{6} $ & $ 642 $ & $ 181 $ & $ 165 $ & $ 3.2 \times 10^{17} $ \\ 
\hline 
$ 0.25 $ & $ 0.3 $ & $ 1 $ & $ 3.5 $ & $ 0.012 $ & $ 1.5 \times 10^{-11} $ & $ 10^{5} $ & $ 287 $ & $ 81 $ & $ 37 $ & $ 2.9 \times 10^{16} $ \\ 
$ 0.25 $ & $ 0.091 $ & $ 1 $ & $ 3.5 $ & $ 0.012 $ & $ 1.5 \times 10^{-10} $ & $ 10^{4} $ & $ 90 $ & $ 25 $ & $ 11 $ & $ 9.2 \times 10^{14} $ \\ 
$ 0.25 $ & $ 0.9 $ & $ 1 $ & $ 3.5 $ & $ 0.012 $ & $ 1.5 \times 10^{-12} $ & $ 10^{6} $ & $ 908 $ & $ 256 $ & $ 117 $ & $ 9.2 \times 10^{17} $ \\ 
$ 0.25 $ & $ 1.3 $ & $ 1 $ & $ 3.5 $ & $ 0.012 $ & $ 7.4 \times 10^{-13} $ & $ 2.0 \times 10^{6} $ & $ 1284 $ & $ 362 $ & $ 165 $ & $ 2.6 \times 10^{18} $ \\ 
\hline
$ 0.025 $ & $ 0.3 $ & $ 1 $ & $ 3.5 $ & $ 1.2 \times 10^{-3} $ & $ 1.5 \times 10^{-11} $ & $ 10^{4} $ & $ 287 $ & $ 81 $ & $ 11 $ & $ 2.9 \times 10^{16} $ \\ 
$ 0.025 $ & $ 0.9 $ & $ 1 $ & $ 3.5 $ & $ 1.2 \times 10^{-3} $ & $ 1.5 \times 10^{-12} $ & $ 10^{5} $ & $ 908 $ & $ 256 $ & $ 37 $ & $ 9.2 \times 10^{17} $ \\ 
$ 0.025 $ & $ 2.9 $ & $ 1 $ & $ 3.5 $ & $ 1.2 \times 10^{-3} $ & $ 1.5 \times 10^{-13} $ & $ 10^{6} $ & $ 2873 $ & $ 810 $ & $ 117 $ & $ 2.9 \times 10^{19} $ \\ 
$ 0.025 $ & $ 4.1 $ & $ 1 $ & $ 3.5 $ & $ 1.2 \times 10^{-3} $ & $ 7.4 \times 10^{-14} $ & $ 2.0 \times 10^{6} $ & $ 4063 $ & $ 1146 $ & $ 165 $ & $ 8.2 \times 10^{19} $ \\ 
\hline
$ 2.5 \times 10^{-3} $ & $ 0.9 $ & $ 1 $ & $ 3.5 $ & $ 1.2 \times 10^{-4} $ & $ 1.5 \times 10^{-12} $ & $ 10^{4} $ & $ 908 $ & $ 256 $ & $ 11 $ & $ 9.2 \times 10^{17} $ \\ 
$ 2.5 \times 10^{-3} $ & $ 2.9 $ & $ 1 $ & $ 3.5 $ & $ 1.2 \times 10^{-4} $ & $ 1.5 \times 10^{-13} $ & $ 10^{5} $ & $ 2873 $ & $ 810 $ & $ 37 $ & $ 2.9 \times 10^{19} $ \\ 
$ 2.5 \times 10^{-3} $ & $ 9.1 $ & $ 1 $ & $ 3.5 $ & $ 1.2 \times 10^{-4} $ & $ 1.5 \times 10^{-14} $ & $ 10^{6} $ & $ 9085 $ & $ 2562 $ & $ 117 $ & $ 9.2 \times 10^{20} $ \\ 
$ 2.5 \times 10^{-3} $ & $ 12 $ & $ 1 $ & $ 3.5 $ & $ 1.2 \times 10^{-4} $ & $ 7.4 \times 10^{-15} $ & $ 2.0 \times 10^{6} $ & $ 1.3 \times 10^{4} $ & $ 3624 $ & $ 165 $ & $ 2.6 \times 10^{21} $ \\ 
\hline 
\end{tabular}
\end{table*}

\begin{table*}[h!]
\caption{Possible scaling for the different plasma $\beta$ simulations using constant atmospheric temperature $T_0$ (in K) and time $t_0$ (in s), while varying the volume magnetic field $B_0$ (in G). The other quantities are: length $L_0$ (in Mm), pressure $P_0$ (in Pa), density $\rho_0$ (in$\U{kg\ m^{-3}}$), velocity $V_0$ (in$\U{km\ s^{-1}}$), Alfv\'en speed $c_A$ (in$\U{km\ s^{-1}}$), sound speed $c_S$ (in$\U{km\ s^{-1}}$), and energy $E_0$ (in J).
}
\label{Tab:BetascalingB} 
\centering
\begin{tabular}{ccccccccccc}
\hline 
$\beta$ & $L_0 $ & $t_0$ & $B_0$ & $P_0$ & $\rho_0$ & $T_0$ & $V_0$ & $c_A$ & $c_S$ & $E_0$ \\
\hline
$ 1.0 $ & $ 0.5 $ & $ 1 $ & $ 0.5 $ & $ 10^{-3} $ & $ 1.2 \times 10^{-13} $ & $ 10^{6} $ & $ 454 $ & $ 128 $ & $ 117 $ & $ 2.3 \times 10^{15} $ \\
$ 1.0 $ & $ 0.5 $ & $ 1 $ & $ 1 $ & $ 4.0 \times 10^{-3} $ & $ 4.8 \times 10^{-13} $ & $ 10^{6} $ & $ 454 $ & $ 128 $ & $ 117 $ & $ 9.4 \times 10^{15} $ \\ 
$ 1.0 $ & $ 0.5 $ & $ 1 $ & $ 5 $ & $ 0.1 $ & $ 1.2 \times 10^{-11} $ & $ 10^{6} $ & $ 454 $ & $ 128 $ & $ 117 $ & $ 2.3 \times 10^{17} $ \\
\hline 
$ 0.25 $ & $ 0.9 $ & $ 1 $ & $ 0.5 $ & $ 2.5 \times 10^{-4} $ & $ 3.0 \times 10^{-14} $ & $ 10^{6} $ & $ 908 $ & $ 256 $ & $ 117 $ & $ 1.9 \times 10^{16} $ \\ 
$ 0.25 $ & $ 0.9 $ & $ 1 $ & $ 1 $ & $ 10^{-3} $ & $ 1.2 \times 10^{-13} $ & $ 10^{6} $ & $ 908 $ & $ 256 $ & $ 117 $ & $ 7.5 \times 10^{16} $ \\ 
$ 0.25 $ & $ 0.9 $ & $ 1 $ & $ 5 $ & $ 0.025 $ & $ 3.0 \times 10^{-12} $ & $ 10^{6} $ & $ 908 $ & $ 256 $ & $ 117 $ & $ 1.9 \times 10^{18} $ \\
$ 0.25 $ & $ 0.9 $ & $ 1 $ & $ 10 $ & $ 0.1 $ & $ 1.2 \times 10^{-11} $ & $ 10^{6} $ & $ 908 $ & $ 256 $ & $ 117 $ & $ 7.5 \times 10^{18} $ \\ 
\hline 
$ 0.025 $ & $ 2.9 $ & $ 1 $ & $ 1 $ & $ 10^{-4} $ & $ 1.2 \times 10^{-14} $ & $ 10^{6} $ & $ 2873 $ & $ 810 $ & $ 117 $ & $ 2.4 \times 10^{18} $ \\ 
$ 0.025 $ & $ 2.9 $ & $ 1 $ & $ 5 $ & $ 2.5 \times 10^{-3} $ & $ 3.0 \times 10^{-13} $ & $ 10^{6} $ & $ 2873 $ & $ 810 $ & $ 117 $ & $ 5.9 \times 10^{19} $ \\
$ 0.025 $ & $ 2.9 $ & $ 1 $ & $ 10 $ & $ 0.01 $ & $ 1.2 \times 10^{-12} $ & $ 10^{6} $ & $ 2873 $ & $ 810 $ & $ 117 $ & $ 2.4 \times 10^{20} $ \\ 
$ 0.025 $ & $ 2.9 $ & $ 1 $ & $ 100 $ & $ 1 $ & $ 1.2 \times 10^{-10} $ & $ 10^{6} $ & $ 2873 $ & $ 810 $ & $ 117 $ & $ 2.4 \times 10^{22} $ \\ 
\hline 
$ 2.5 \times 10^{-3} $ & $ 9.1 $ & $ 1 $ & $ 5 $ & $ 2.5 \times 10^{-4} $ & $ 3.0 \times 10^{-14} $ & $ 10^{6} $ & $ 9085 $ & $ 2562 $ & $ 117 $ & $ 1.9 \times 10^{21} $ \\ 
$ 2.5 \times 10^{-3} $ & $ 9.1 $ & $ 1 $ & $ 10 $ & $ 10^{-3} $ & $ 1.2 \times 10^{-13} $ & $ 10^{6} $ & $ 9085 $ & $ 2562 $ & $ 117 $ & $ 7.5 \times 10^{21} $ \\ 
$ 2.5 \times 10^{-3} $ & $ 9.1 $ & $ 1 $ & $ 100 $ & $ 0.1 $ & $ 1.2 \times 10^{-11} $ & $ 10^{6} $ & $ 9085 $ & $ 2562 $ & $ 117 $ & $ 7.5 \times 10^{23} $ \\ 
\hline 
\end{tabular}
\end{table*}

\begin{table*}[h!]
\caption{Possible scaling for the different plasma $\beta$ simulations using constant volume magnetic field $B_0$ (in G) and length $L_0$ (in Mm), while varying the atmospheric temperature $T_0$ (in K). The other quantities are: time $t_0$ (in s), pressure $P_0$ (in Pa), density $\rho_0$ (in$\U{kg\ m^{-3}}$), velocity $V_0$ (in$\U{km\ s^{-1}}$), Alfv\'en speed $c_A$ (in$\U{km\ s^{-1}}$), sound speed $c_S$ (in$\U{km\ s^{-1}}$), and energy $E_0$ (in J).
}
\label{Tab:BetascalingB2} 
\centering
\begin{tabular}{ccccccccccc}
\hline 
$\beta$ & $L_0 $ & $t_0$ & $B_0$ & $P_0$ & $\rho_0$ & $T_0$ & $V_0$ & $c_A$ & $c_S$ & $E_0$ \\
\hline
$ 1.0 $ & $ 5 $ & $ 110 $ & $ 5 $ & $ 0.1 $ & $ 1.2 \times 10^{-9} $ & $ 10^{4} $ & $ 45 $ & $ 12 $ & $ 11 $ & $ 3.1 \times 10^{20} $ \\ 
$ 1.0 $ & $ 5 $ & $ 34 $ & $ 5 $ & $ 0.1 $ & $ 1.2 \times 10^{-10} $ & $ 10^{5} $ & $ 143 $ & $ 40 $ & $ 37 $ & $ 3.1 \times 10^{20} $ \\ 
$ 1.0 $ & $ 5 $ & $ 11 $ & $ 5 $ & $ 0.1 $ & $ 1.2 \times 10^{-11} $ & $ 10^{6} $ & $ 454 $ & $ 128 $ & $ 117 $ & $ 3.1 \times 10^{20} $ \\ 
$ 1.0 $ & $ 5 $ & $ 3.5 $ & $ 5 $ & $ 0.1 $ & $ 1.2 \times 10^{-12} $ & $ 10^{7} $ & $ 1436 $ & $ 405 $ & $ 370 $ & $ 3.1 \times 10^{20} $ \\
\hline
$ 0.25 $ & $ 5 $ & $ 55 $ & $ 5 $ & $ 0.025 $ & $ 3.0 \times 10^{-10} $ & $ 10^{4} $ & $ 90 $ & $ 25 $ & $ 11 $ & $ 3.1 \times 10^{20} $ \\ 
$ 0.25 $ & $ 5 $ & $ 17 $ & $ 5 $ & $ 0.025 $ & $ 3.0 \times 10^{-11} $ & $ 10^{5} $ & $ 287 $ & $ 81 $ & $ 37 $ & $ 3.1 \times 10^{20} $ \\ 
$ 0.25 $ & $ 5 $ & $ 5.5 $ & $ 5 $ & $ 0.025 $ & $ 3.0 \times 10^{-12} $ & $ 10^{6} $ & $ 908 $ & $ 256 $ & $ 117 $ & $ 3.1 \times 10^{20} $ \\ 
$ 0.25 $ & $ 5 $ & $ 1.7 $ & $ 5 $ & $ 0.025 $ & $ 3.0 \times 10^{-13} $ & $ 10^{7} $ & $ 2873 $ & $ 810 $ & $ 370 $ & $ 3.1 \times 10^{20} $ \\ \hline 
$ 0.025 $ & $ 5 $ & $ 17 $ & $ 5 $ & $ 2.5 \times 10^{-3} $ & $ 3.0 \times 10^{-11} $ & $ 10^{4} $ & $ 287 $ & $ 81 $ & $ 11 $ & $ 3.1 \times 10^{20} $ \\ 
$ 0.025 $ & $ 5 $ & $ 5.5 $ & $ 5 $ & $ 2.5 \times 10^{-3} $ & $ 3.0 \times 10^{-12} $ & $ 10^{5} $ & $ 908 $ & $ 256 $ & $ 37 $ & $ 3.1 \times 10^{20} $ \\ 
$ 0.025 $ & $ 5 $ & $ 1.7 $ & $ 5 $ & $ 2.5 \times 10^{-3} $ & $ 3.0 \times 10^{-13} $ & $ 10^{6} $ & $ 2873 $ & $ 810 $ & $ 117 $ & $ 3.1 \times 10^{20} $ \\ 
$ 0.025 $ & $ 5 $ & $ 0.6 $ & $ 5 $ & $ 2.5 \times 10^{-3} $ & $ 3.0 \times 10^{-14} $ & $ 10^{7} $ & $ 9085 $ & $ 2562 $ & $ 370 $ & $ 3.1 \times 10^{20} $ \\ 
\hline 
$ 2.5 \times 10^{-3} $ & $ 5 $ & $ 5.5 $ & $ 5 $ & $ 2.5 \times 10^{-4} $ & $ 3.0 \times 10^{-12} $ & $ 10^{4} $ & $ 908 $ & $ 256 $ & $ 11 $ & $ 3.1 \times 10^{20} $ \\ 
$ 2.5 \times 10^{-3} $ & $ 5 $ & $ 1.7 $ & $ 5 $ & $ 2.5 \times 10^{-4} $ & $ 3.0 \times 10^{-13} $ & $ 10^{5} $ & $ 2873 $ & $ 810 $ & $ 37 $ & $ 3.1 \times 10^{20} $ \\ 
$ 2.5 \times 10^{-3} $ & $ 5$ & $ 0.6 $ & $ 5 $ & $ 2.5 \times 10^{-4} $ & $ 3.0 \times 10^{-14} $ & $ 10^{6} $ & $ 9085 $ & $ 2562 $ & $ 117 $ & $ 3.1 \times 10^{20} $ \\ 
$ 2.5 \times 10^{-3} $ & $ 5 $ & $ 0.2 $ & $ 5 $ & $ 2.5 \times 10^{-4} $ & $ 3.0 \times 10^{-15} $ & $ 10^{7} $ & $ 2.9 \times 10^{4} $ & $ 8104 $ & $ 370 $ & $ 3.1 \times 10^{20} $ \\ 

\hline 
\end{tabular}
\end{table*}

\begin{table*}[h!]
\caption{Possible scaling for the different plasma $\beta$ simulations using constant atmospheric pressure $P_0$ (in Pa), while varying the temperature $T_0$ (in K) and length $L_0$ (in Mm). The other quantities are: time $t_0$ (in s), volume magnetic field $B_0$ (in G), pressure $P_0$ (in Pa), density $\rho_0$ (in$\U{kg\ m^{-3}}$), velocity $V_0$ (in$\U{km\ s^{-1}}$), Alfv\'en speed $c_A$ (in$\U{km\ s^{-1}}$), sound speed $c_S$ (in$\U{km\ s^{-1}}$), and energy $E_0$ (in J).
}
\label{Tab:BetascalingL} 
\centering
\begin{tabular}{ccccccccccc}
\hline 
$\beta$ & $L_0 $ & $t_0$ & $B_0$ & $P_0$ & $\rho_0$ & $T_0$ & $V_0$ & $c_A$ & $c_S$ & $E_0$ \\
\hline
$ 1.0 $ & $ 0.01 $ & $ 0.2 $ & $ 0.5 $ & $ 10^{-3} $ & $ 1.2 \times 10^{-11} $ & $ 10^{4} $ & $ 45 $ & $ 12 $ & $ 11 $ & $ 2.5 \times 10^{10} $ \\ 
$ 1.0 $ & $ 0.05 $ & $ 1.1 $ & $ 0.5 $ & $ 10^{-3} $ & $ 1.2 \times 10^{-11} $ & $ 10^{4} $ & $ 45 $ & $ 12 $ & $ 11 $ & $ 3.1 \times 10^{12} $ \\ 
$ 1.0 $ & $ 0.1 $ & $ 2.2 $ & $ 0.5 $ & $ 10^{-3} $ & $ 1.2 \times 10^{-11} $ & $ 10^{4} $ & $ 45 $ & $ 12 $ & $ 11 $ & $ 2.5 \times 10^{13} $ \\ 
$ 1.0 $ & $ 0.5 $ & $ 11 $ & $ 0.5 $ & $ 10^{-3} $ & $ 1.2 \times 10^{-11} $ & $ 10^{4} $ & $ 45 $ & $ 12 $ & $ 11 $ & $ 3.1 \times 10^{15} $ \\ 
\hline
$ 0.25 $ & $ 0.1 $ & $ 0.2 $ & $ 1 $ & $ 10^{-3} $ & $ 4.0 \times 10^{-13} $ & $ 3.0 \times 10^{5} $ & $ 497 $ & $ 140 $ & $ 64 $ & $ 10^{14} $ \\ 
$ 0.25 $ & $ 0.5 $ & $ 1.0 $ & $ 1 $ & $ 10^{-3} $ & $ 4.0 \times 10^{-13} $ & $ 3.0 \times 10^{5} $ & $ 497 $ & $ 140 $ & $ 64 $ & $ 1.2 \times 10^{16} $ \\ 
$ 0.25 $ & $ 1 $ & $ 2.0 $ & $ 1 $ & $ 10^{-3} $ & $ 4.0 \times 10^{-13} $ & $ 3.0 \times 10^{5} $ & $ 497 $ & $ 140 $ & $ 64 $ & $ 1.0 \times 10^{17} $ \\ 
$ 0.25 $ & $ 5 $ & $ 10 $ & $ 1 $ & $ 10^{-3} $ & $ 4.0 \times 10^{-13} $ & $ 3.0 \times 10^{5} $ & $ 497 $ & $ 140 $ & $ 64 $ & $ 1.3 \times 10^{19} $ \\ 
\hline
$ 0.025 $ & $ 1 $ & $ 0.3 $ & $ 3.2 $ & $ 10^{-3} $ & $ 1.2 \times 10^{-13} $ & $ 10^{6} $ & $ 2873 $ & $ 810 $ & $ 117 $ & $ 1.0 \times 10^{18} $ \\ 
$ 0.025 $ & $ 5 $ & $ 1.7 $ & $ 3.2 $ & $ 10^{-3} $ & $ 1.2 \times 10^{-13} $ & $ 10^{6} $ & $ 2873 $ & $ 810 $ & $ 117 $ & $ 1.3 \times 10^{20} $ \\ 
$ 0.025 $ & $ 10 $ & $ 3.5 $ & $ 3.2 $ & $ 10^{-3} $ & $ 1.2 \times 10^{-13} $ & $ 10^{6} $ & $ 2873 $ & $ 810 $ & $ 117 $ & $ 10^{21} $ \\ 
$ 0.025 $ & $ 20 $ & $ 7.0 $ & $ 3.2 $ & $ 10^{-3} $ & $ 1.2 \times 10^{-13} $ & $ 10^{6} $ & $ 2873 $ & $ 810 $ & $ 117 $ & $ 8.0 \times 10^{21} $ \\\hline 
$ 2.5 \times 10^{-3} $ & $ 1 $ & $ 0.064 $ & $ 10 $ & $ 10^{-3} $ & $ 4.0 \times 10^{-14} $ & $ 3.0 \times 10^{6} $ & $ 1.6 \times 10^{4} $ & $ 4439 $ & $ 203 $ & $ 1.0 \times 10^{19} $ \\ 
$ 2.5 \times 10^{-3} $ & $ 5 $ & $ 0.3 $ & $ 10 $ & $ 10^{-3} $ & $ 4.0 \times 10^{-14} $ & $ 3.0 \times 10^{6} $ & $ 1.6 \times 10^{4} $ & $ 4439 $ & $ 203 $ & $ 1.2 \times 10^{21} $ \\ 
$ 2.5 \times 10^{-3} $ & $ 10 $ & $ 0.6 $ & $ 10 $ & $ 10^{-3} $ & $ 4.0 \times 10^{-14} $ & $ 3.0 \times 10^{6} $ & $ 1.6 \times 10^{4} $ & $ 4439 $ & $ 203 $ & $ 10^{22} $ \\ 
$ 2.5 \times 10^{-3} $ & $ 20 $ & $ 1.3 $ & $ 10 $ & $ 10^{-3} $ & $ 4.0 \times 10^{-14} $ & $ 3.0 \times 10^{6} $ & $ 1.6 \times 10^{4} $ & $ 4439 $ & $ 203 $ & $ 8.0 \times 10^{22} $ \\ 
\hline 
\end{tabular}
\end{table*}

\end{document}